\documentclass[twocolumn]{article}
\usepackage[margin=0.55in]{geometry}
\usepackage[T1]{fontenc}
\usepackage[utf8]{inputenc}
\usepackage{lmodern}
\usepackage[english]{babel}
\usepackage{soul}
\usepackage[table]{xcolor}
\sethlcolor{yellow}
\usepackage{amssymb}
\usepackage{pifont}%

\usepackage{multirow,float}

\usepackage{graphicx}

\usepackage{hyperref}
\hypersetup{
    colorlinks=true,
    linkcolor=black,
    citecolor=black,
    filecolor=black,
    urlcolor=black,
    linkbordercolor = {1 1 1},
}

\usepackage[square,super,numbers,sort&compress]{natbib}
\title{Crowdsourcing Bridge Vital Signs with Smartphone Vehicle Trips}
\author{
    Thomas J. Matarazzo$^{1,2,3,\ast}$, D\'aniel Kondor$^4$,  Sebastiano Milardo$^1$, Soheil S. Eshkevari$^{1,5}$,
    \\ Paolo Santi$^{1,6}$, Shamim N. Pakzad$^5$, Markus J. Buehler$^7$, and Carlo Ratti$^{1,2}$,  \\
    \normalsize{$^1$Senseable City Laboratory, MIT, Cambridge MA 02139 USA}\\
    \normalsize{$^2$Cornell Tech, Cornell University, New York, NY 10044 USA}\\
    \normalsize{$^3$United States Military Academy, West Point, NY 10996 USA}\\
	\normalsize{$^4$Singapore-MIT Alliance for Research and Technology, Singapore}\\
	\normalsize{$^5$Dept. of Civil and Environmental Engineering, Lehigh University, Bethlehem, PA 18015 USA}\\
	\normalsize{$^6$Istituto di Informatica e Telematica del CNR, Pisa, Italy}\\
	\normalsize{$^7$Dept. of Civil and Environmental Engineering, Massachusetts Institute of Technology, Cambridge, MA 02139 USA}\\
	\normalsize{$^\ast$ E-mail: \texttt{tomjmat@mit.edu}}
}
\date{\today}

\begin{document}

\maketitle

\subsection*{Abstract}
A key challenge in monitoring and managing the structural health of bridges is the high-cost associated with specialized sensor networks. In the past decade, researchers predicted that cheap, ubiquitous mobile sensors would revolutionize infrastructure maintenance; yet many of the challenges in extracting useful information in the field with sufficient precision remain unsolved. Herein it is shown that critical physical properties, e.g., modal frequencies, of real bridges can be determined accurately from everyday vehicle trip data.  The primary study collects smartphone data from controlled field experiments and ``uncontrolled'' UBER rides on a long-span suspension bridge in the USA and develops an analytical method to accurately recover modal properties. The method is successfully applied to ``partially-controlled'' crowdsourced data collected on a short-span highway bridge in Italy. This study verifies that pre-existing mobile sensor data sets, originally captured for other purposes, e.g., commercial use, public works, etc., can contain important structural information and therefore can be repurposed for large-scale infrastructure monitoring. A supplementary analysis projects that the inclusion of crowdsourced data in a maintenance plan for a new bridge can add over fourteen years of service ($30\%$ increase) without additional costs. These results suggest that massive and inexpensive datasets collected by smartphones could play an important role in monitoring the health of existing transportation infrastructure.

\subsection*{Introduction}
Mobile sensors could reform the way we measure infrastructure health. Smartphones contain dozens of sensors that are carried by almost 50\% of the population globally\cite{pewsmartphone}. Analyses of crowdsourcing networks have uncovered truths about the social, economical, civil, and technological systems we rely on in an urban environment, e.g., quantifying urban human mobility\cite{gonzalez2008understanding, Schneider2013}, understanding the perception of built environment\cite{Li2019}, modeling and predicting infectious disease spread\cite{Massaro2019}, etc. \cite{Pelletier2011,bbva,hawelka2014geo}. Recently, there has been an increased focus on self-sustaining sensing platforms such as data generated by vehicle fleets, either with smartphones or dedicated sensors\cite{Massaro2017,anjomshoaa2018city,OKeeffe12752}.

The effectiveness of a crowdsourcing application is a question of precision and scale. While crowdsourced data has a proven value and cost-effectiveness in a variety of large-scale applications, there remain fundamental challenges for those that call for precise measurements in time and space. Civil infrastructure monitoring techniques require highly curated data, often sampled at a high rate by synchronous data-acquisition systems and low noise sensors. Over the last decade, researchers have been eager to validate crowdsourcing broadly to apply to existing infrastructure; however, spatiotemporal precision challenges have limited advancements in related work to synthetic models and idealized experiments \mbox{ \cite{yang2004extracting,lin2005use,yang2009extracting,siringoringo2012estimating,zhang2012damage,feng2015citizen,mcgetrick2017implementation, yang2020vehicle, sitton2020bridge, sitton2020frequency}}.

There is a global need to endorse infrastructure monitoring to optimize the service of the most critical assets of highway networks and the urban environment. A failure to maintain a city’s bridges, buildings, and other infrastructure can have disastrous consequences, from enormous repair costs to the loss of human life \cite{furtado2014cost}. The vast number of U.S. bridges with structural problems accentuates shortcomings in bridge maintenance protocols\cite{ASCE2017,ARBTA2019}. Modern bridge condition assessments are based on field inspection notes from visual inspections rather than large digital datasets; a paradigm that severely limits the frequency of structural health assessments, the depth of the information collected, and the ability to execute preventive maintenance. 

Crowdsourcing bridge vibration data would modernize structural health monitoring (SHM) and bridge asset management at a global scale. Longitudinal data collection and analyses are essential for tracking changes in structural state, informing preemptive repairs, and service life analyses \cite{pandey1994damage,peeters2001one,kong2003evaluation, yang2004hilbert,nair2006time, cross2013long,yao2012autoregressive}. In typical SHM applications, a synchronized sensor network is mounted on a bridge\footnote{scale on the order of hectometers to kilometers} to measure acceleration\footnote{scale on the order of millimeters per second squared (milli-\it{G})}. While sensor data provides advantages over field inspections, due to high costs, such static sensor networks are rarely incorporated in a bridge management system. Despite the implications of Moore's Law, installation and maintenance costs are often still too high for the vast majority of bridge owners.

\begin{figure*}[t]
\centering
\includegraphics[width=\textwidth]{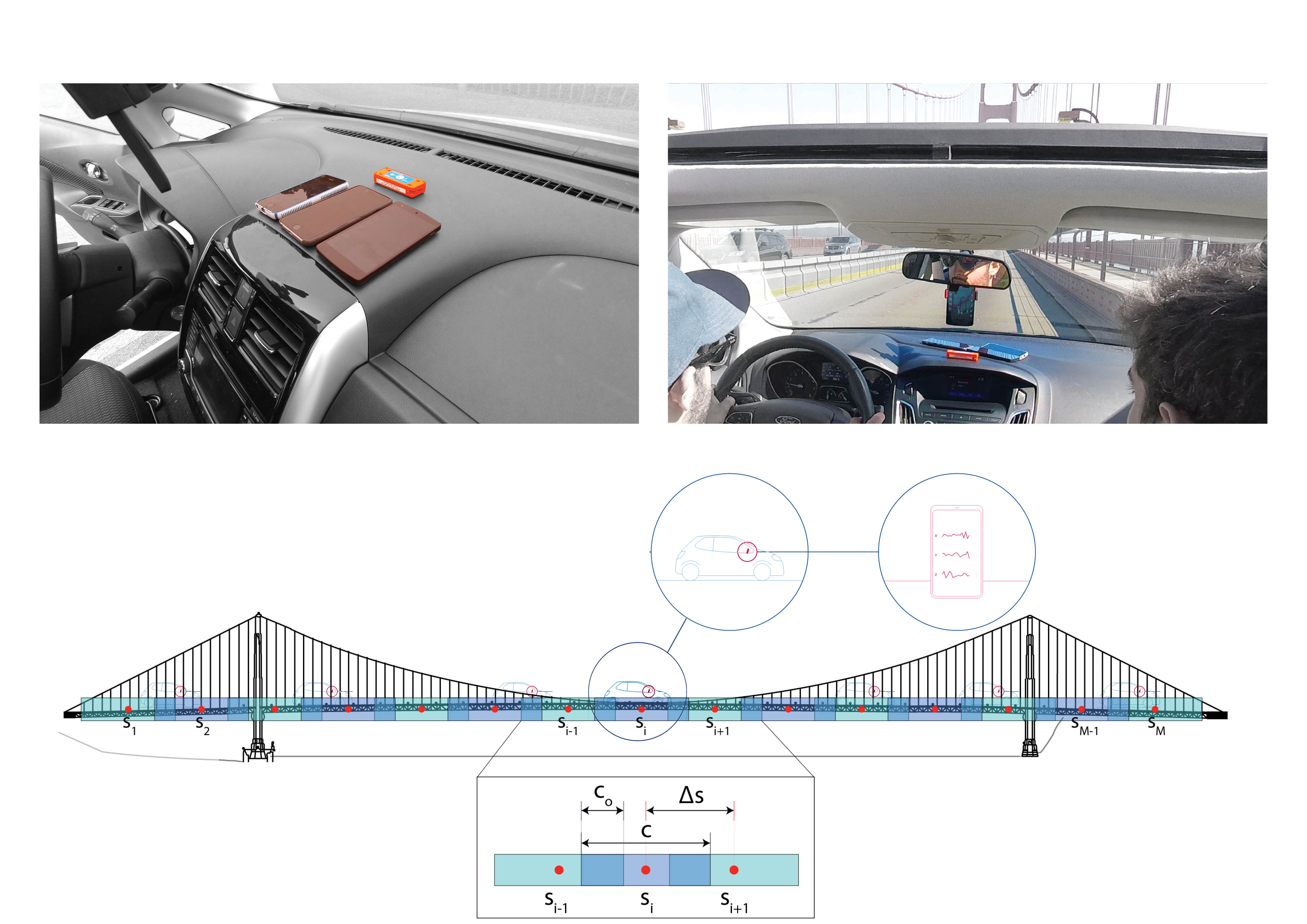}
\caption{Illustration of data collection and the spatial segmentation used in this study. Top: photographs of one of the 102 ``controlled'' vehicle trips over the Golden Gate Bridge displaying the sensor layout on the dashboard of one of the vehicles (Ford Focus). The smartphones were facing upward, such that one axis was well-aligned with gravity. Such an orientation is not strictly necessary; although, knowledge on the configuration of the sensors is helpful for data preprocessing. Bottom: generic schematic of spatial segmentation of a bridge which is defined through two independent parameters: $\Delta s$ and $c$, which remain uniform over the length of the bridge. The red circles represent the centers of each segment, while the light colored boxes show the segment widths. A close-up of three adjacent segments $s_{i-1}$, $s_i$, and $s_{i+1}$, is shown to detail the segmentation parameters: $c$ is the length of each segment, $c_o$ is the length of the overlap between segments, and $\Delta s$ is the distance between the centers (red circles) of adjacent segments.}
\label{fig:SpatialGrouping}
\end{figure*}

Mobile sensor networks resolve this financial bottleneck by bypassing high-end sensing systems. Smartphones or other cheap sensors, either mounted on \cite{anjomshoaa2018city} or `riding in’ vehicles \cite{OKeeffe12752}, can contribute useful data. The primary appeal of a mobile sensor network is that it does not require dedicated devices: it can repurpose existing ones resulting in cheaper and more convenient data collection compared to traditional methods. For example, smartphones can scan a city’s infrastructure with a wide spatiotemporal coverage at little or no cost through the existing travel patterns of the ``host vehicles'', e.g., taxis  \cite{OKeeffe12752}. A recent study showed that just two mobile sensors produces SHM information comparable to 240 static sensors\cite{matarazzo2018scalable}, with other studies reporting similarly large efficiencies \cite{marulanda2016modal}. Simultaneously, applications of deep learning in material science have exemplified how sparse structural response measurements can be used to accurately predict global behavior, e.g., load distributions \cite{guo2020semi}, and optimize multiscale design.

Nevertheless, there remain questions on estimation precision in a real setting, for which a clear, successful SHM application of mobile sensors is lacking. Studies considering smartphone data from vehicles driving over bridges in typical scenarios \cite{mcgetrick2017implementation,matarazzo2018crowdsensing} have been unable to extract explicit, meaningful information. No prior work has considered data from smartphones in moving automobiles and successfully determined modal properties of a real bridge \mbox{ \cite{shirzad2020frequency,mei2020towards,elhattab2019extraction,matarazzo2018crowdsensing,mcgetrick2017implementation,ozer2015citizen,siringoringo2012estimating,zhang2012damage,yang2009extracting,lin2005use,yang2004extracting,sitton2020bridge,sitton2020frequency}}. Likewise, no prior work has considered uncontrolled or partially-controlled smartphone data (from ridesourcing or crowdsourcing), i.e., data for which the analysts had little or no influence on key data collection parameters (see \emph{Methods} for detailed classifications of data controllability).

As such, the core question remains unanswered: \textit{can crowdsourcing produce precise structural health information under real-world conditions?} This paper provides strong evidence that supports crowdsourcing. It shows that data collected by smartphones in moving vehicles under real-world conditions can be used to identify structural modal properties of a bridge, information which is vital to condition assessments and damage detection frameworks \cite{farrar1996damage,wahab1999damage,peeters2001one,yang2004extracting,nair2006time,yao2012autoregressive,cross2013long}. The following results based on ``uncontrolled'' and ``partially-controlled'' crowdsensed data collected on two structurally diverse bridges, verify that pre-existing mobile sensing mechanisms and data sets, originally established for other purposes, can produce important structural information and therefore could be utilized to extensively monitor the health of \emph{networks} of bridges, worldwide.

\begin{figure*}[h]
\centering
\includegraphics[width=\textwidth]{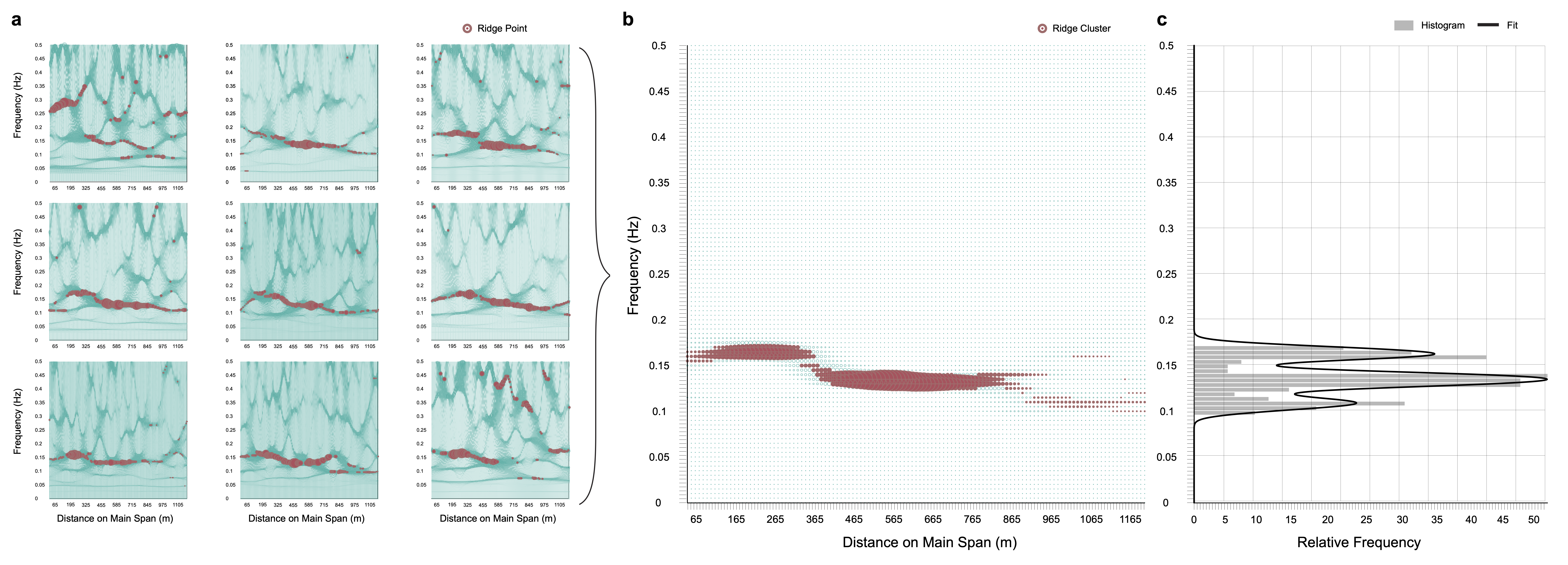}
\caption{Illustration of the main methodology used to extract most probable modal frequencies (MPMFs). (a) The synchrosqueezed wavelet transform is calculated for each of the bride crossings individually. The time variable is then remapped into linear location ($r$) on the bridge, resulting in the space-frequency representation of the signal. Ridges are then identified as peaks at each location. (Steps 1-6 in the methodology shown in Table~\ref{tab:methods}). (b) Peaks from each individual location are first aggregated in the spatial groups shown in Fig.~\ref{fig:SpatialGrouping}, then among all datasets, resulting in one space-frequency diagram of identified ridge clusters. Each location in space represents one spatial segment from Fig.~\ref{fig:SpatialGrouping} (Step 7). (c) The most prominent vibration frequencies from each spatial group are selected and a histogram of these is created. The modes of this histogram are identified using a kernel density (KDE) fit; these picks are considered the MPMFs (Steps 8-9 in Table~\ref{tab:methods}).}
\label{fig:wsst_max}
\end{figure*}

\subsection*{Results}
\paragraph{Determining most probable modal frequencies.} The Golden Gate Bridge is a long-span suspension bridge in California, USA, with a 1280-meter-long main span. Two distinct datasets were collected: a field study where researchers drove over the bridge 102 times, recording data with two smartphones (iPhone 5 and iPhone 6) and data collected by UBER drivers in 72 bridge trips during normal operations. In the following, these are referred to as ``controlled'' and ``ridesourcing'' datasets. Impressively, while these datasets are relatively small, they produce accurate results, demonstrating that mobile-sensor-based SHM can be applied easily, cheaply, and immediately in the real world.

Figure~\ref{fig:SpatialGrouping} describes the controlled data collection process and the spatial analysis method. The experiments focus on the identification of the first ten vertical and torsional frequencies of the Golden Gate Bridge, which are below 0.5 Hz (see \emph{Methods} for a complete catalog). 

The methodology developed to determine the \emph{most probable modal frequencies} (MPMFs), the main result of the analysis, is illustrated in Figure~\ref{fig:wsst_max}. The plots in Figure~\ref{fig:wsst_max}(a) are produced via the synchrosqueezed wavelet transform \cite{daubechies2011synchrosqueezed} (with Morlet basis) and a mapping of time to a local coordinate system (in space) based on simultaneous GPS measurements. Figure~\ref{fig:wsst_max}(b) shows the data aggregation step: a plot of the frequencies that were most consistently present over all the trips versus bridge length (space). In the final step, depicted in Fig. ~\ref{fig:wsst_max}(c), a kernel density estimate (KDE) is fit to the histogram (displayed as solid line) of the frequency candidates. Further details are presented in the \emph{Methods} section and \emph{Supplementary Material}.

\begin{figure*}[!h]
\centering
\includegraphics[width=\textwidth]{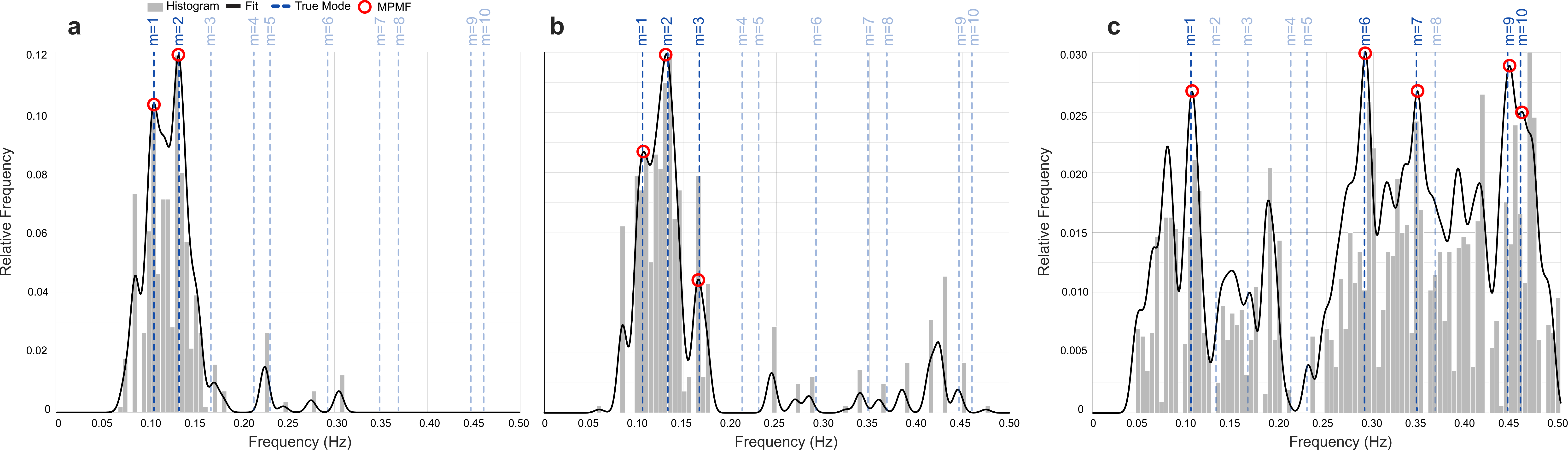}

\caption{Final estimates of probability density functions (PDFs) from controlled trips from (a)~iPhone~5 and (b)~iPhone~6. A histogram (grey bars) and kernel density estimates (black line, Gaussian kernel with bandwidth of 0.005 Hz) is shown for each dataset. Blue dashed lines refer to true modal frequencies of the bridge (see Table~\ref{tab:MPMFs}). The local maxima (modes) of each multimodal PDF corresponds to a possible modal frequency. The iPhone 5 MPMFs correspond to the two largest peaks, which estimate the first and second modal frequencies (m=1,2) . The iPhone 6 MPMFs correspond to the three largest peaks, which estimate the first, second, and third modal frequencies (m=1,2,3). The known modal frequencies of the Golden Gate Bridge are indicated for reference. Details of the MPMFs displayed in this plot are provided in Table~\ref{tab:MPMFs}. (c) PDFs based on ridesourcing data. Since there are over a dozen MPMF candidates (local maxima), a significance threshold of $0.90$ was set for the CDF values, which resulted in five MPMFs. Details of the MPMFs displayed in this plot are provided in Table~\ref{tab:UberMPMFs}.}
\label{fig:hist1}
\end{figure*}

The MPMFs are defined as the peaks of the KDE probability density function. The MPMF results for the controlled data and ridesourcing data and their corresponding probability density functions are displayed in Fig.~\ref{fig:hist1}; these plots highlight the likely vibrational frequencies. The MPMF values are displayed in Tables~\ref{tab:MPMFs} and~\ref{tab:UberMPMFs} for each dataset, respectively. For KDE, a Gaussian kernel was selected with a bandwidth equal to 1\% of the frequency range considered (0.005 Hz since the range is 0-0.5Hz). Initially, MPMFs were chosen by visual inspection of the PDF. The corresponding cumulative distribution function (CDF) indicated that the chosen MPMFs for the iPhone 5 were in the upper $2\%$ and those MPMFs for the iPhone 6 were in the upper $10\%$ (see Table~\ref{tab:MPMFs} for precise values). That is for all MPMFs in this work, the selection was automated by setting an upper threshold for the CDF of the frequency candidates, e.g., $10\%$.

The MPMF results are compared with the the most comprehensive report on the modal properties of the Golden Gate Bridge. The true values in Tables~\ref{tab:MPMFs} and~\ref{tab:UberMPMFs} are based on data collected over a three-month period with a wireless network of 240 accelerometers and found over sixty vibrational modes (vertical, transverse, and torsional)~\citep{pakzad2008design, pakzad2009statistical}. In particular, there are seven vertical modes and three torsional modes below 0.5 Hz (see Table~\ref{tab:GGBfreq}). The fundamental vibrational modes (lowest frequencies) are often the most significant contributors to the overall dynamic response of a structure \citep{chopra2001dynamics}; i.e., they are most important.

Overall, the first two modal frequencies of the bridge are estimated accurately by both the iPhone 5 and the iPhone 6 in our controlled experiments (see Fig.~\ref{fig:hist1} and Table~\ref{tab:MPMFs}). The iPhone 5 data estimates the first frequency as 0.106 Hz and the second frequency as 0.132 Hz. Similarly, the iPhone 6 data estimates first frequency as 0.108 Hz and the second frequency as 0.132 Hz. In addition, the iPhone 6 data estimates the third modal frequency as 0.166 Hz. Estimates made by the iPhone 5 for both frequencies and the iPhone 6 for the second frequency are accurate up to three significant digits, the precision used in our study ($< 0.5\%$ error). Estimates of the first and third frequency by the iPhone 6 have errors of 1.9\% and 2.3\% respectively. These errors are low and within an acceptable range in the context of operational modal identification. In practice, independent methods are employed to corroborate estimates of modal properties. For a given dataset, modal frequency estimates will vary based on the method yet will often be within $1-2\%$ of each other \cite{matarazzo2016stride}; in some cases discrepancies can be as high as $5-7\%$ for certain modes, methods, and datasets \cite{siringoringo2012estimating}.

The ridesourcing data consists of 72 data sets representing 37  different vehicles and 19 smartphone models, which were generally collected at higher driving speeds (see Tables~ \ref{tab:vehicles}, \ref{tab:phones}, and Figure ~\ref{fig:uber_speeds} in \emph{Methods}). In PDF from the ridesourcing data, there are over a dozen MPMF candidate peaks that are distributed more evenly (see Fig.~\ref{fig:hist1}c). The MPMFs were chosen as the peaks in the upper $10\%$ of the CDF, which resulted in five candidates. Table ~\ref{tab:UberMPMFs} compares these MPMFs with the true modal frequencies. Impressively, each of the five MPMFs corresponds to a true modal frequency and includes the fundamental mode ($m = 1$). Additionally, the MPMFs here include four new modes that were not detected in the controlled trip data ($m = 6, 7, 9, 10$); this reflects how the content of the frequency information is subject to the context of the vehicle trips, such as, vehicle attributes, smartphone sensors, etc. In general, in order to be measured, the frequency content must be present in the bridge vibrations, meaning the modes of interest must be externally excited by the dynamic loads, e.g., traffic, wind, etc. (see \emph{Supplementary Material} for further details).

\begin{table}
  \caption{Most probable modal frequencies (MPMFs) for the controlled data, extracted from the peaks of Fig ~\ref{fig:hist1}a and Fig ~\ref{fig:hist1}b. The vibration mode numbers are indicated by $m$, where frequencies are sorted in ascending order. MPMFs (estimates) and true frequency values are in Hertz with corresponding errors (\%) in parentheses. CDF values indicate the significance of each peak in the overall PDF.}
\label{tab:MPMFs}
  \centering

\begin{tabular}{cccccc}
    \hline
    &
    \multicolumn{2}{c}{iPhone 5}&
    \multicolumn{2}{c}{iPhone 6}\\
    \cline{2-5}
    $m$ & MPMF & CDF & MPMF & CDF & True \\
    \hline 
    1   &   0.106 (0.000) &   0.98  &  0.108  (1.89)& 0.95 & 0.106\\
    2   &   0.132 (0.000) &   1.00  &  0.132  (0.000) & 1.00 & 0.132\\
    3   &   --     &   --    &  0.166 (2.35)  & 0.90 & 0.170\\
    \hline
\end{tabular}
\end{table}

The fundamental vertical frequency ($m = 1$) and the third torsional frequency ($m = 9$) were estimated perfectly to the nearest thousandth ($0.000\%$ error). The fifth vertical frequency ($m = 6$), the second torsional frequency ($m = 7$),  and the seventh vertical frequency ($m = 10$), were estimated with errors of $3.3\%$, $2.3\%$, and $0.65\%$, respectively. It is important to note that the proposed method can detect both vertical and torsional modes because the dataset provided vertical (gravity direction) acceleration measurements. In this study, distinctions between vertical and torsional MPMFs were made by matching them with the closest known modal frequencies (see Tables ~\ref{tab:UberMPMFs} and ~\ref{tab:GGBfreq}). The inclusion of other acceleration channels may enable the detection of longitudinal and transverse modes.

There are more peaks in the PDF from the ridesourcing data than there are in those from the controlled data. Furthermore, many of the peaks in the ridesourcing PDF appear relatively significant, whereas in the other PDFs, there were $2-3$ primary peaks and the others were negligible. This may suggest that there is a higher potential for false positives when dealing with ``uncontrolled'' datasets. On the other hand, this may be an artifact of a relatively small sample size that could be solved with larger volumes of data.

\begin{table}
\caption{MPMFs for the ridesourcing data, were extracted as the peaks of Fig ~\ref{fig:hist1}c in the top ten percentile (five in total). The vibration mode numbers are indicated by $m$, where frequencies are sorted in ascending order. MPMFs (estimates) and true frequency values are in Hertz with corresponding errors (\%) in parentheses. CDF values indicate the significance of each peak in the overall PDF}
\label{tab:UberMPMFs}
\centering

\begin{tabular}{cccc}
    \hline
    $m$ & MPMF & CDF & True\\
    \hline 
    1   &   0.106  (0.000) &   0.97  & 0.106\\
    6   &   0.291  (3.32)&   1.00  & 0.301 \\
    7   &   0.347  (2.35)&   0.97  & 0.339 \\
    9   &   0.445  (0.000) &   0.99  & 0.445 \\
    10   &  0.458  (0.650)&   0.93  & 0.461 \\
    \hline
    \end{tabular}
    \end{table}

\begin{figure}[t]
    \centering
    \includegraphics[width=\columnwidth]{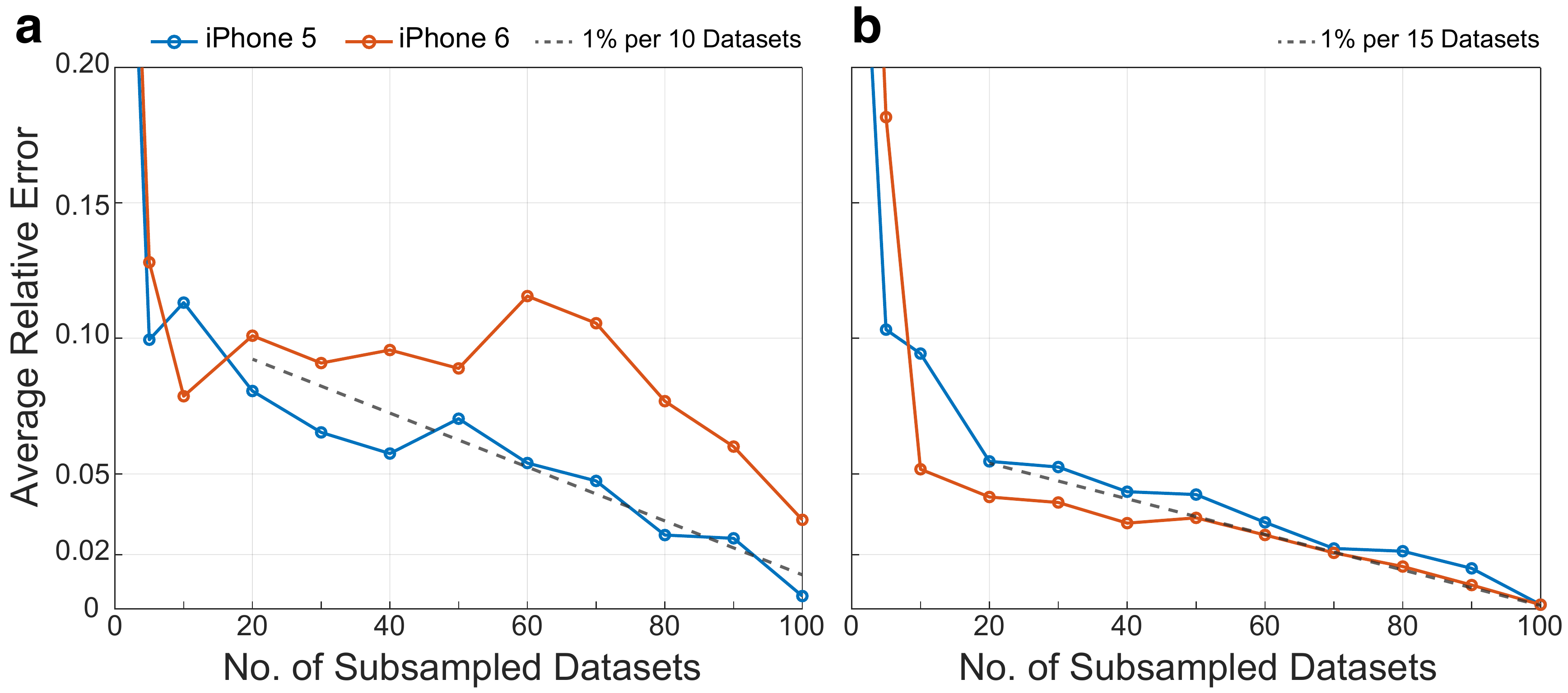}
    \caption{Errors of the MPMFs as a function of the size of the data subset: a) smartphone MPMF errors for the first vertical frequency ($m=1$); b) smartphone MPMF errors for the second vertical frequency ($m=2$). Each line represents the mean error for a different smartphone.}
    \label{fig:IndErr}
\end{figure}

\paragraph{Number of vehicle trips versus accuracy.}
The quantity of bridge trips in both tests is arbitrary: one could reasonably ask \textit{``How many datasets are needed to accurately estimate a bridge modal frequency?''}. Answers to questions such as this could help produce practical guidelines that drive large-scale efforts to regularly collect vehicle scanning data. This was considered using random subsets of the controlled trips, i.e., choosing $N_S < 102$ bridge trips and repeating the analysis to extract MPMFs. The following analysis focuses on the first two modal frequencies as those were detected by both sensors previously. The average detection error (i.e.~difference between MPMFs and true values) is provided in Fig.~\ref{fig:IndErr}. Overall, these curves show that the frequency extraction procedure excels as more data becomes available.

Fig.~\ref{fig:IndErr}a shows the behavior of the first frequency errors as data subsets grow larger and Fig.~\ref{fig:IndErr}b shows the same relationship for the second frequency. In all cases, the MPMF errors fall to the order of 10\% when the number of datasets reaches 10. Beyond this point, a gradual (not purely monotonic) increase in accuracy continues as the dataset size increases towards $100$. With only 40 datasets, one can see that the iPhone 5 estimates for both frequencies become quite accurate; within 6\% of the true values, which is also true for both smartphone estimates of the second modal frequency. 

Overall, errors for the second frequency decreased more consistently and rapidly than those for the first frequency. In other words, less effort was required to estimate the second modal frequency with a high accuracy. With either smartphone, an estimate of the second modal frequency can be achieved within 5\% of the true value with as few as 30 datasets; and with 50 datasets, the error reduces to 4\%. 

These plots indicate that, depending on the smartphone sensor, only a relatively small amount of datasets is needed to get a rough estimate of a modal frequency -- between 10 and 50 datasets can achieve an error on the order of 10\%; however, a considerably larger amount of data, about 80 or 90 datasets, is needed in order to reduce errors to the order of 3\%. More specifically, once the error falls below 10\% error, each additional 10 datasets tends to reduce it further by about 1\%. 

\begin{figure*}[!h]
\centering
\includegraphics[width=0.85\textwidth]{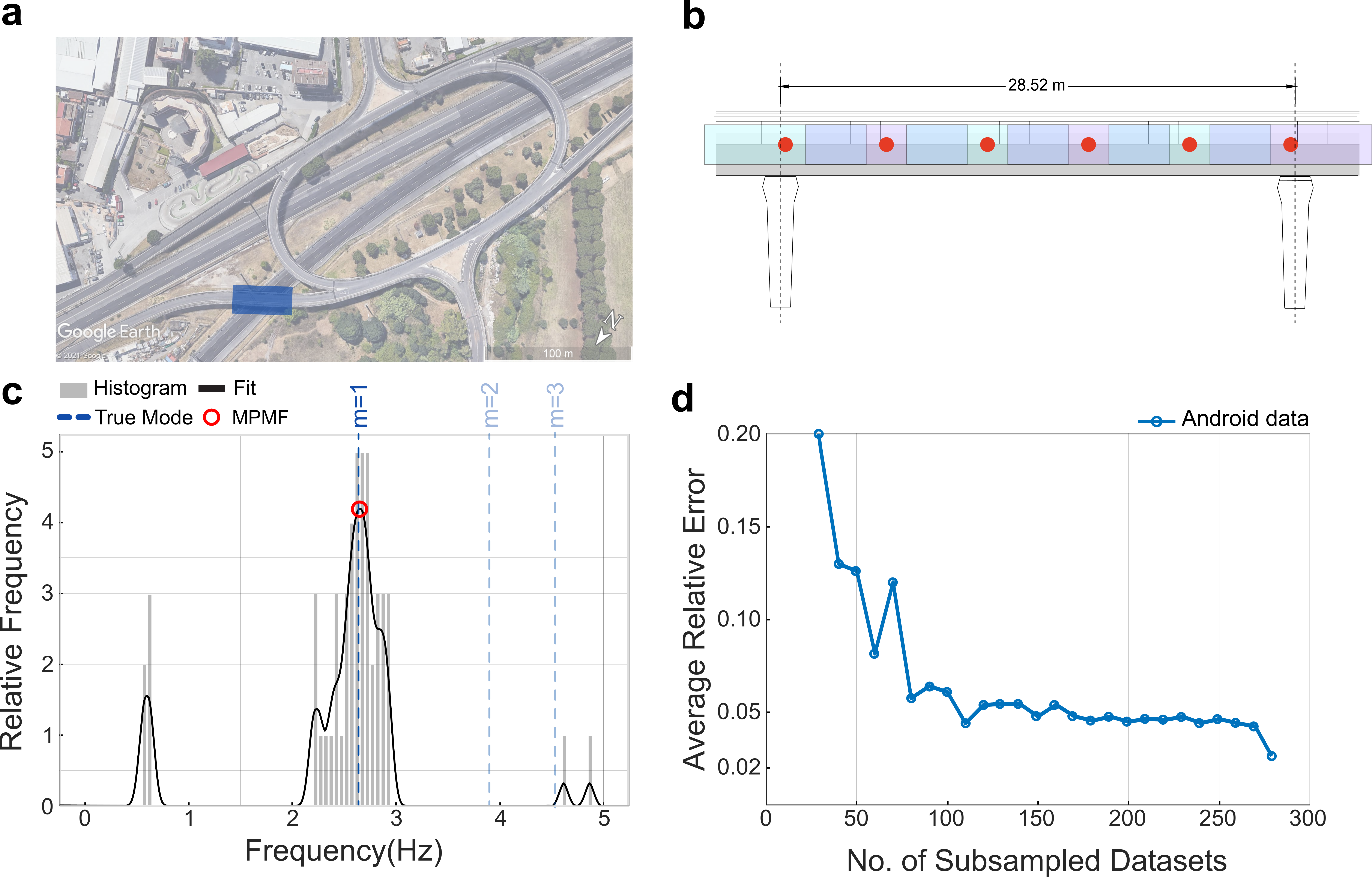}
\caption{Summary of the crowdsensing application to the short-span concrete bridge, one of the most common types of bridges in the US. (a) an aerial view of the E80 highway exchange in Ciampino, Italy (image from \emph{Google Earth}) with the studied bridge span highlighted in blue (b) the geometry and spatial segmentation of the bridge (c) the histogram and the fitted probability density function found from processing 280 smartphone datasets. The probability density function indicates a large peak at 2.64 Hz which is near the bridge's fundamental frequency ($m=1$) and is the only MPMF. The reference frequencies for the first three modes are shown with blue vertical dotted lines (d) the average error per dataset. The error reduces substantially as the number of scans increases. Overall, the error reduces more rapidly for $0<N<100$ than for $N>100$.}
\label{fig:Ciampino}
\end{figure*}

\paragraph{Application on short highway bridge.}
The Golden Gate Bridge is an atypical bridge. Suspension bridges comprise less than $1 \% $ of US bridges. They have flexible main spans that usually exceed 500 m in length, corresponding to low fundamental modal frequencies (on the order of tenths of Hz). The proposed method was applied to a 28-meter-long reinforced concrete bridge to evaluate its performance on a short span, which better represents a typical highway bridge. According to the FHWA national bridge inventory, $41 \%$ of bridges in the US are concrete bridges with spans shorter than 30 m \cite{fhwa}. It is difficult to establish expected values for the fundamental frequencies of such bridges; however, they can be considered greater than $1$ Hz \cite{en19912, mohseni2014simplified} and are therefore more likely to coincide with vehicle modal frequencies, which usually range $1-3$ Hz \cite{olley1934independent,crolla1999olley}.

The studied bridge is in Ciampino, Italy and is part of the European route E80 in the Rome metropolitan area monitored by ANAS Sp.A. The highway exchange is a system of ramps and turns, consisting of 30 short bridge spans(see Fig.~\ref{fig:Ciampino}a). The bridge span was instrumented with six accelerometers to record ambient vibrations nearly continuously from September 2020 to April 2021 (see Fig.~\ref{fig:Ciampino_fixed} in \emph{Methods} for the sensor layout). Successive batches of data (collected in two-hour intervals) were processed using the automated frequency domain decomposition (AFDD) algorithm \cite{brincker2007automated} to determine the structural modal properties of the bridge and track them over time (hours, days, weeks). Based on these analyses, the fundamental frequency ($m=1$) of the bridge of interest (see Fig.~\ref{fig:Ciampino_fixed}) was determined as 2.58 Hz; this is the reference value for the mobile sensing results.

Smartphone-vehicle trips were recorded using an Android-based smartphone application that was developed by the researchers for the ANAS Sp.A. road maintenance crew; as such the data was ``partially-controlled'' by the analysts (see Table \ref{tab:controlability} in  \emph{Methods}). The app was equipped with a geofence to automatically trigger data collection whenever the crew traveled over the bridge span of interest. Over 250 bridge trips have been collected by the crew using mobile devices so far. The proposed method processed 280 datasets to estimate MPMFs and compare with the identified fundamental frequency. Figures \ref{fig:Ciampino}b-d  illustrate the spatial segmentation pattern, the PDF of the frequency candidates using Gaussian KDE, and the error versus number of trips. As shown in Fig.~\ref{fig:Ciampino}c, the MPMF from mobile sensor data matches the reference frequency ($2.32 \%$ error). The PDF also shows some small, statistically insignificant peaks at other frequencies. The second and third frequencies were not detected in this batch of data. As observed when comparing ``controlled'' and ``ridesourcing'' datasets, the observability of modes depends on many physical factors such as bridge excitation, vehicle attributes, etc. Note vehicle-bridge-road interaction (VBRI) effects are not accounted for in the proposed method. Recent work has shown how information about the vehicle system, e.g., transfer function, and road profile can improve the accuracy of the estimated bridge modal properties \cite{gonzalez2012identification, mcgetrick2015experimental, eshkevari2019bridge, eshkevari2020bridge} (see \emph{Supplementary Material} for discussion).

Figure~ \ref{fig:Ciampino}d, presents the effect of sample size on the fundamental frequency estimation error. The analysis conducted was identical to that which produced Figure~ \ref{fig:IndErr} (based on random subsets of data). As the number of samples increases to 100, the error drops to $5\%$. Beyond this point, the error continues to decline, at a much slower rate. The result with the smallest error ($2.32\%$) is that which uses all 280 datasets.

There are three main observations:
First, similar to the error trends seen for the long-span bridge in Figure \ref{fig:IndErr}, in this application, the errors decrease in two phases, a rapid change followed by a gradual one. The difference being the approximate $N$ value at which this transition occurs ($N \approx 20$ vs. $N \approx 100$). Second, the error of the fundamental frequency estimate at $N=100$ datasets is $5.5\%$, which is comparatively higher than the primary application (see Figure~ \ref{fig:IndErr}). Part of this difference may be attributed to the uncontrollability of the data or the fact that trips on short-span bridges will generally contain fewer data samples due to short-time durations. Future work may utilize metadata on the uncontrolled factors such as smartphone mount, or vehicle  model, to quantify estimation uncertainty. In addition, specialized methods for sensor noise reduction may be useful \cite{elhattab2019extraction}. Finally, the spatial data aggregation technique was effective despite relatively imprecise GPS data. Reported GPS errors were about 4.3 m on average, which represented about $15\%$ of the length of the bridge (see \emph{Methods} for details). Analyses of noisy acceleration data in the \emph{Supplementary Material} show similar results. Overall, the successful applications suggest that the proposed method may apply to a broad range of sensor qualities and bridge types.

\paragraph{Discussion.} This paper proves that bridge vibration frequencies can be identified from smartphone-vehicle trip data in real-world conditions. It is emphasized that data from a single trip is insufficient; yet as few as 100 crowdsourced datasets can produce useful modal frequency estimates (below 6 \% error) for both short-span and long-span bridges. The number of trips used in the primary study was less than $0.1\%$ of the daily trips made on the Golden Gate Bridge; this shows that there is an enormous sensing potential represented by smartphones globally that contains valuable information about bridges and other important infrastructure. Furthermore, the accuracy of the most-probable modal frequencies (MPMFs) improved as the number of datasets increased.

Through a broad set of successful real-world applications, this paper strongly demonstrates the robustness of crowdsourced smartphone data for bridge health monitoring. The results exemplified three broad classes of crowdsourced data from which modal properties can be accurately extracted: ``controlled'', ``uncontrolled'', and ``partially-controlled''. This shows that analysts do not necessarily need to design or influence data collection features such as vehicle speed, smartphone orientation, etc., in order for the datasets to provide value to a bridge management system. Pre-existing mobile sensor datasets, originally captured for other purposes, e.g., ridesourcing, public works, etc., may be repurposed for infrastructure monitoring.

The MPMF analyses did not explicitly account for vehicle-bridge-road-interaction (VBRI) effects yet were still successful in identifying bridge modal frequencies accurately. This highlights the strong and persistent traces of bridge modal frequencies among the collected datasets in varied vehicle and road conditions. Future applications that are more sensitive to VBRI features may benefit from incorporating smartphone metadata, e.g., smartphone mount, vehicle model, roadway type, etc., and vehicle dynamical properties to achieve precise MPMFs or to quantify uncertainties with regard to data controllability. In short, entropy is a desirable feature of crowdsourced vehicle-trip data which will help reduce bias in estimated bridge properties (see \emph{Supplementary Material} for a detailed discussion). Recommended future work can quantify relationships between data controllability parameters and the accuracy of modal property estimates.

Long-term studies on bridge dynamic features over various environmental and operational conditions establish a foundation for the rapid development and verification of new methods for data-driven condition evaluations.  Frequencies are a gateway to additional modal properties and structural features which have explicit relationships with certain types of bridge damage and deterioration, e.g., mode shape curvature, local stiffness, etc.  \mbox{\cite{pandey1994damage,farrar1996damage,wahab1999damage,ruzzene1997natural,peeters2001one,gonzalez2012identification}}. Historically, such techniques have been confined to academic research applications, which have only impacted a small fraction of existing bridges \mbox{\cite{lee2005neural, sahin2003quantification,  ralbovsky2010frequency, kim2007vibration, fan2011vibration,peeters2001one, jin2016damage, liang2018frequency}}. Furthermore, the majority of these studies are based on short-term data (a few months); long-term structural behavior (at least one year of data) is essential to establish the baselines and performance benchmarks needed for condition evaluations \mbox{\cite{smith2016studies}}. 

It is important to emphasize that "machine intelligence", e.g., a trained AI model, was not needed to determine accurate MPMF results from aggregated smartphone-vehicle trip data. The real power of crowdsourced data lies in the ease of obtaining longitudinal data, which fuels a nearly continuous monitoring, feature extraction, and detection system with which a present state can be analyzed and compared to historical baselines (past seasons over years). With these unprecedented data volumes, automated structural condition assessment systems would no longer be confined to methods rooted in surrogate model optimization \mbox{\cite{catbas2007limitations, schlune2009improved, xiao2015multiscale}}; they could finally, truly benefit from advances in artificial intelligence and computational learning as seen in fields such as image and video recognition, recommendation systems, or multimodal sentiment analysis  \mbox{\cite{he2016deep,krizhevsky2012imagenet,brunton2016discovering,vaswani2017attention,raissi2018hidden,brock2018large}}. When fueled with long-term monitoring data, artificial intelligence has the potential to provide bridge engineers and owners with unprecedented information for maintenance and operation at virtually little to no extra cost.

The potential benefits of this paradigm are quantified in analyses (see \emph{Supplementary Material}) which generate reliability profiles to simulate how bridge service life is affected by the availability of condition information in a maintenance plan. The analyses predicted that crowdsourced monitoring accumulates information that adds over two years of service to a 43-year-old bridge ($15\%$ increase) and adds almost fifteen years of service to a brand new bridge ($30 \%$ increase). The findings of this paper could have an immediate impact on the management of bridges in many countries; they emphasize the benefit of integrating crowdsourced monitoring information into a bridge management plan as soon as the bridge is in operation.

\paragraph{Acknowledgements}
The authors would like to thank Anas S.p.A., Allianz, Brose, Cisco, Dover Corporation, Ford, the Amsterdam Institute for Advanced Metropolitan Solutions, the Fraunhofer Institute, the Kuwait-MIT Center for Natural Resources and the Environment, Lab Campus, RATP, Singapore--MIT Alliance for Research and Technology (SMART), SNCF Gares \& Connexions, UBER, the U.S. Department of Defense High-Performance Computing Modernization Program, and all the members of the MIT Senseable City Lab Consortium for supporting this research. The authors would like to thank Tom Benson for his help with Figures 1, 2, 3, and 5.
The authors would also like to thank Dr. Kai Guo for providing constructive feedback on this work. 

\paragraph{Author contributions}
T.J.M., P.S., and C.R. designed the research; T.J.M., D.K., S.M., S.S.E., P.S., S.N.P., and C.R. performed research; T.J.M., D.K., S.M., and S.S.E. analyzed the data; T.J.M., D.K., S.M., S.S.E., P.S., S.N.P., and M.B. wrote the paper. The authors declare no conflict of interest.

\normalsize

\section*{Methods}

\subsection*{Data Controllability Definitions}

\begin{table*}
  \caption{Definition of data controllability and comparison: What level of influence did the analysts have on the following data collection parameters? (\ding{51} = "Complete Control", \ding{53} = "No Control" , \ding{46} = "Guidance Provided, No Enforcement")}
  \label{tab:controlability}
  \center
  \begin{tabular}{cccc}
  \hline

    Parameter & Experiment & Ridesourcing & Highway Bridge \\

      & (Controlled) & (Uncontrolled) & (Partially Controlled) \\
    \hline

    Smartphone Model & \ding{51} &  \ding{53}  &  \ding{53}  \\

    Smartphone Mount Type & \ding{51} & \ding{53}  &  \ding{46} \\

    Smartphone Orientation & \ding{51} & \ding{53}  &  \ding{46} \\
 
    Smartphone Sampling Rate & \ding{51} & \ding{53} &  \ding{51}  \\

    Vehicle Make and Model & \ding{51} & \ding{53}  &  \ding{53} \\

    Vehicle Speed & \ding{51} & \ding{53}  &  \ding{46} \\

    Vehicle Route & \ding{51} & \ding{53}  &  \ding{53} \\
    \hline
\end{tabular}
\end{table*}

The data considered in this paper represent three broad classes of crowdsourced data: ``controlled'', ``uncontrolled'', and ``partially-controlled''. Seven controllability dimensions were considered to characterize the smartphone-vehicle trip data: smartphone model, smartphone mount type, smartphone orientation, smartphone sampling rate, vehicle make and model, vehicle speed, and vehicle route. In its simplest terms, controllability metrics address the following question \emph{``What level of influence did the analysts have on the following data collection parameters?''}. For each dimension, there are three possible categories: "Complete Control", "No Control", and ``Guidance Provided, No Enforcement''. ``Complete control'' means the analysts dictated this dimension, while ``No Control'' means the data collector controlled this dimension. ``Guidance Provided, No Enforcement'' means the analysts provided a formal recommendation to the data collector; however, the recommendation was not monitored or enforced whatsoever.

Controlled data is defined as a dataset for which analysts had complete control over all seven controllability dimensions. Uncontrolled data is defined as a dataset in which analysts had no control over any of the seven controllability dimensions. Partially-controlled data is data that is neither controlled nor uncontrolled.Table \ref{tab:controlability} provides details on the controllability of each dataset. 

In summary, the experimental data collected on the Golden Gate Bridge was ``controlled'', the ridesourcing data collected on the Golden Gate Bridge was ``uncontrolled'', and the data collected on the highway bridge in Italy was ``partially-controlled''.

\subsection*{Descriptions of Smartphone-Vehicle Trip Data}

\subsubsection*{Controlled Data}

\begin{table}
  \caption{Ten modal frequencies of the Golden Gate Bridge from \cite{pakzad2009statistical}: seven vertical and three torsional, all below 0.5 Hz. The vibration mode numbers are indicated by $m$, where frequencies are sorted in ascending order; mode types ``V'' and ``T'' refer to vertical and torsional modes; ``A'' and ``S'' refer to anti-symmetric and symmetric mode shapes with respect to the middle of the main bridge span.}
  \label{tab:GGBfreq}
  \centering
  \begin{tabular}{ccccl}
  \hline
    $m$ & Type & Shape & Frequency (Hz)\\
    \hline
    1 & V & A & 0.106  \\
    2 & V & S & 0.132 \\
    3 & V & S & 0.170 \\
    4 & V & A & 0.216 \\
    5 & T & A & 0.230 \\
    6 & V & S & 0.301 \\
    7 & T & S & 0.339 \\
    8 & V & A & 0.371 \\
    9 & T & A & 0.445 \\
    10 & V & S & 0.461 \\
    \hline
\end{tabular}
\end{table}

The controlled data were recorded by an iPhone 5 and iPhone 6 using the \emph{Sensor Play} App which recorded about one dozen data fields including sensor measurements, e.g., triaxial acceleration, GPS, gyroscope, magnetometer, etc., and inferred values, e.g., speed, GPS error. The trip data was collected in morning and afternoon rush-hour periods in five consecutive days (June 18 - 22, 2017). In the raw data,  triaxial acceleration was sampled at about $100 $ Hz and GPS coordinates were sampled at about $1$ Hz. The acceleration data was irregularly sampled. Temporal jitter was observed in most of the individual samples, i.e., non-uniform sampling periods. Additionally, when comparing sampling rates of independent data sets (different trips), the average actual sampling rate varied about $1-5\%$ from the target value (the data was often undersampled). Therefore each acceleration dataset was resampled to $100$ Hz before processing. Since the positions and orientations of the sensors were known, the acceleration channel used in the analysis was the one which coincided with gravity. In summary, the measurement channels used to determine MPMFs were triaxial acceleration and GPS.

Vehicle trips in the controlled dataset were completed during morning and afternoon rush hour periods. For each trip, the driver and passenger qualitatively rated the traffic levels on a $1-3$ scale, where ``1'' means no congestion and ``3'' means ``stop-and-go'' conditions. Overall, these ratings were not significant, and only helped to roughly correlate traffic with vehicle speed. Nonetheless, it is important to track traffic volumes, e.g., car counts, as it directly influences the amplitudes of the bridge vibrations.

As a form of controlled variety in the trips, two sedan-style vehicles were used and five target speeds were defined: 32, 40, 48, 56, and 64 kph.\footnote{The speed limit on the bridge is 72 kph; however, there are 40 kph advisories for areas near the toll gates, located at the south end of the bridge.}. The vehicle speed was monitored visually by the driver (cruise control was not used). Instantaneous speed data (automatically produced through the app) was used to compare with the speed targets. Overall, the median speed values matched well with the target speeds. The first fifty trips were completed using a Nissan Sentra and the remaining fifty-two trips used a Ford Focus. Each vehicle trip was assigned a predetermined target speed; while crossing over the bridge spans, the driver almost always\footnote{During one northbound trip, the driver was ordered by an authority to increase the speed of the vehicle; this resulted in an average speed well above the target speed of 48 kph. During one southbound trip, significant congestion resulted in an average speed that was well below the target speed of 32 kph. Two extra trips were made to account for these instances, hence 102 in total.} maintained an average speed within $\pm$4 kph of the target. A summary of vehicle trip details, e.g., the trip speed distribution is provided in Table~\ref{tab:trips}.

\begin{table}[b]
  \caption{Details of the 102 vehicle trips made over the Golden Gate Bridge. Trips were completed using two different vehicles driving with five constant speeds. The distribution of the trips taken for each speed is provided. The majority of the trips were at 40 kph (29 out of 102) and the minority of the trips were at 64 kph (10 out of 102). Traffic ratings were recorded qualitatively on a scale of 1-3 for each trip based on observations of the researchers. A rating of ``3'' indicates very high traffic, e.g.,``bumper to bumper''.}
  \label{tab:trips}
  \center
  \begin{tabular}{ccc}
  \hline
    Target Avg. Speed &No. of Trips &Avg. Traffic Rating\\
    \hline
    32 & 26 &  1.5  \\
    40 & 29 & 1.2 \\
    48 & 17& 1.2 \\
    56 & 20 & 1.1 \\
    64 & 10 & 1.0 \\
    \hline
\end{tabular}
\end{table}

\subsubsection*{Uncontrolled Ridesourcing Data}
Vehicle trips supplied by the ridesourcing operator, UBER, were made by 37 distinct vehicle types that together constitute a set of typical vehicle makes and models used in such fleets. The largest group, 9 trips were made by Toyota Priuses; a breakdown of the number of trips by vehicle type is presented in Table ~\ref{tab:vehicles}. Measurements were collected by the ridesourcing app itself during regular operations in its default commercial-use settings, i.e., no modifications were made to accommodate bridge vibration analysis. The analysts had no control over the data prior to its collection, i.e., no control over parameters listed in able~ \ref{tab:controlability}. The ridesourcing operator, UBER, provided data as requested, exclusively based on a geographical area of interest (the Golden Gate Bridge), without any other specifications.

Each dataset (trip) contained about one dozen data fields, which included sensor measurements, e.g., triaxial acceleration, GPS, gyroscope, magnetometer, etc., inferred values, e.g., speed, GPS error, and vehicle and smartphone model information. The triaxial acceleration data was provided in a raw, unknown orientation, which required correction in order to extract the channel corresponding to the vertical (gravity) direction. The Nericell approach ~\cite{mohan2008nericell} was selected to reorient the acceleration signals. This method uses triaxial accelerations and GPS to estimate tilting angles and reorients them based on Euler angles. The average GPS error value was about 7.7 m. In summary, the analysis to determine MPMFs used only triaxial acceleration and GPS data.

Drivers used multiple types of smartphones, a summary of which is given in  Table ~\ref{tab:phones}. In total, in 19 trips, the device used was an iPhone model, ranging from iPhone 5s to iPhone X and including multiple model variants. In 51 trips, the device was a Samsung model, with the Galaxy S5 and Note 4 being the most popular (17 and 10 trips respectively). Most trips were driven at relatively high speeds, above $70$ kph (43 mph), while the rest of trip speeds were distributed relatively evenly below that; it was assumed that this resulted from varying traffic conditions on the bridge.

\begin{figure}
    \centering
    \includegraphics{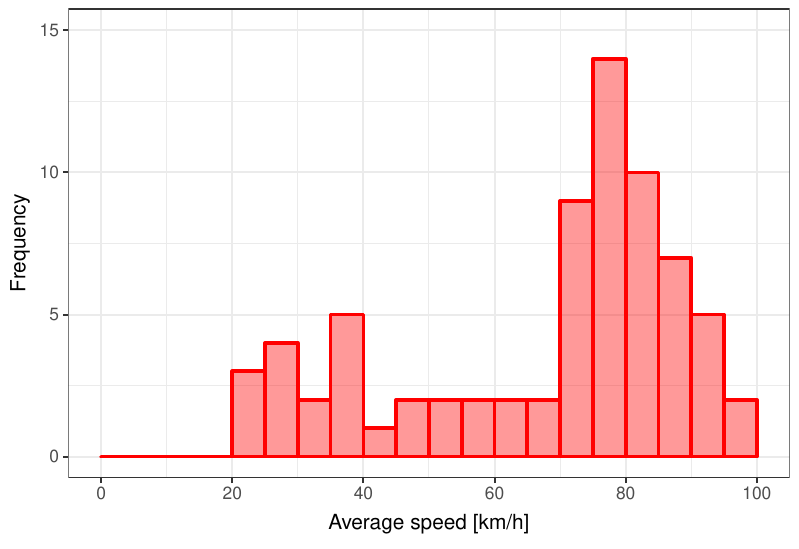}
    \caption{Histogram of average vehicle speeds recorded in the ridesourcing dataset.}
    \label{fig:uber_speeds}
\end{figure}

\subsubsection*{Partially-Controlled Data}

For the short-span highway bridge, 280 smartphone-vehicle trips were recorded using an Android-based smartphone application that was developed by the research team and was used by the ANAS Sp.A. road maintenance crew. The app is equipped with a geofence to automatically trigger the data collection whenever the crew traveled over the bridge of interest. The data collected included direct measurements of triaxial acceleration, GPS,  gyroscope, and inferred values such as orientations (rotation) and GPS error. The rotation vector is a post-processed vector produced by Android that represents the location and orientation of the device. In summary, the provided rotation vector consists of the last three components of the unit quaternion that comprise the necessary information for device reorientation. A comprehensive description of the reorientation method using quaternion vectors is presented in \cite{shoemake1985animating,kuipers1999quaternions}. The Nericell reorientation approach used for the ridesourcing data was not suitable because orientations were already available through the data collection app.  After reorienting the triaxial acceleration signals, the channel corresponding to gravity was selected for further processing.

It is important to note that the Nericell approach relies on GPS measurements which generally have lower SNRs on shorter bridges. The GPS sampling rate for the smartphone used in this research is 1 Hz. The average GPS accuracy reported by the operating system of the device was 4.3 m with regard to a 28-meter-long bridge. The UBER dataset showed noisier GPS data with an average error of 7.7 m, yet the bridge considered was 1280 m long. To preprocess the GPS data, the coordinates are first projected to a 1D axis along with the longitudinal axis of the bridge (see Fig. ~\ref{fig:real_gps}). Next, the data points with GPS coordinates outside predefined bounding boxes were discarded. After removing outliers, the missing coordinates were recalculated by linear interpolation. The GPS errors did not appear to negatively impact MPMF estimation. Overall, the spatial segmentation and aggregation approach (especially with overlaps) helps to mitigate inaccurate spatial coordinates caused by GPS errors. Future work attempting to extract spatial information of bridge vibration, e.g., structural mode shapes, may require additional processing such as advanced GPS filtering, clustering, and/or sensor data fusion for improved location estimation, e.g. the (SLAM) method \cite{leonard1992dynamic}.

In total, nine different phone models were used and vehicle speeds varied from $9-72$ kph (median of 46.8 kph). More than one vehicle was used for data collection; however, the precise number of different automobiles included was not recorded and is unknown to the analysts. The drivers crossed the bridge during their regular commutes; they were not provided any guidance on vehicle routes, smartphone models, vehicle types, or trip quotas. The analysts  recommended drivers to drive at constant speeds and to mount their phones on a solid holder. The analysts had no control over the smartphone model or vehicles used for data collection and this metadata was not provided for each trip.

\begin{table}
    \caption{List of thirty-seven vehicle types present in the ridesourcing dataset}
    \centering
    \begin{tabular}{l|r}
         Vehicle type & Number of trips \\ \hline
BMW 3-series & 1\\
BMW X1 & 1\\
Chevrolet Cruze & 3\\
Chevrolet Malibu & 2\\
Chevrolet Trax & 2\\
Chrysler Pacifica & 1\\
Ford Focus & 1\\
Ford Fusion & 3\\
Ford Taurus & 1\\
Honda Accord & 3\\
Honda Civic & 2\\
Honda CR-V & 1\\
Honda Insight & 1\\
Honda Odyssey & 1\\
Hyundai Elantra & 2\\
Hyundai Ioniq & 1\\
Hyundai Sonata & 3\\
Kia Forte & 2\\
Kia Optima & 1\\
Lexus RX & 1\\
Mazda CX-5 & 3\\
Mazda CX-9 & 1\\
Mazda MAZDA6 & 1\\
Mercedes-Benz C-Class & 1\\
Mercedes-Benz E-Class & 1\\
Mitsubishi Outlander & 1\\
Nissan Altima & 4\\
Nissan Frontier & 1\\
Subaru Impreza & 1\\
Subaru Outback & 1\\
Toyota C-HR & 1\\
Toyota Camry & 5\\
Toyota Corolla & 2\\
Toyota Prius & 9\\
Toyota RAV4 & 1\\
Toyota Sienna & 4\\
Volkswagen Jetta & 2
    \end{tabular}
    \label{tab:vehicles}
\end{table}

\begin{table}[]
    \caption{List of nineteen smartphone types present in the ridesourcing dataset}
    \centering
    \begin{tabular}{l|r}
         Phone model & Number of trips \\ \hline
iPhone 5s	&	2\\
iPhone 6	&	2\\
iPhone 6s Plus	&	2\\
iPhone 7	&	3\\
iPhone 7 Plus	&	7\\
iPhone X	&	2\\
iPhone 8 Plus	&	1\\
Pixel 2	&	1\\
Pixel 2 XL	&	1\\
Galaxy J7	&	2\\
Galaxy Note 4	&	10\\
Galaxy Note 8	&	3\\
Galaxy Note Edge	&	5\\
Galaxy S4	&	1\\
Galaxy S5	&	17\\
Galaxy S6	&	4\\
Galaxy S7	&	3\\
Galaxy S8	&	5\\
Galaxy S9	&	1
    \end{tabular}
    \label{tab:phones}
\end{table}

\subsection*{Space-frequency representation of the signal}
The instantaneous vertical acceleration measurement at location $r$ on the bridge is represented as a linear combination of oscillatory modes plus noise:
\begin{equation}
    x(r,t) = \sum\limits^D_{d=1} A_d(t) \Phi_d (r) \cos(2\pi\omega_d t )+e(t)
\end{equation}
where $A_d(t)$ is a time-dependent amplitude, $\Phi_d (r)$ describes the spatial mode shape and $\omega_d$ is the vibrational frequency of mode $d$, while $e(t)$ represents the noise. Our main goal is to identify the $\omega_d$ frequencies given the signals $x_i(t) \equiv x(r_i(t),t), \, i = 1,2\dots 102$, where the correspondence between bridge coordinates and time, i.e.~the $r_i(t)$ function is given empirically based on the GPS measurements for each bridge crossing in our dataset. Mathematical connection between these entities and bridge vibrations is discussed in more detail in the Supplementary Material. To achieve this, the time-frequency representation of signal $x_i(t)$ is analyzed based on the assumption that the ``effective'' amplitudes, $A^{\mathrm{eff}}_{di} \equiv A_d(t) \Phi_d(r_i(t))$ are slowly varying functions of time. This is essential since contributions of different modes in the signal vary systematically and stochastically based on the mode shapes and the dynamic forces acting on the bridge; along with the presence of noise, this makes simple spectral methods, e.g.~those based on the Fourier analysis of the whole signal unsuitable. The synchrosqueezed wavelet transform ~\cite{daubechies2011synchrosqueezed,thakur2013synchrosqueezing,jiang2017instantaneous} was employed based on its effectiveness at recovering instantaneous frequencies of noisy signals with many harmonic-like components. This way, for signal $x_i(t)$, a complex valued function $T_{x_i} (f,t)$, is obtained, which gives the instantaneous amplitude of the modal frequency $f$ at time $t$ . Then, the time variable is replaced by its corresponding location $r_i(t)$ in bridge coordinates for each trip in the dataset, obtaining $T_{x_i} (f,r)$, the space-frequency representation of our signal. While in theory, $r$ is often considered a continuous variable, in the following, a discretized version is used which corresponds to a signal downsampled to 1 Hz in time.

\subsection*{Spatial Analysis of Bridge Vibrations}

Modal analysis describes how the presence of each structural vibration mode varies over space and time (see SI for detailed formulation). The synchrosqueezed wavelet transform constructs time-frequency representations to quantify how frequency content in the mobile sensor data varies over time. In mobile acceleration data, the recorded signal includes a mixture of the spatial vibrations of the bridge. The spatial analysis approach in this paper consists of two steps: (i) transformation of GPS points to bridge coordinates; and (ii) spatial segmentation of the bridge. This section describes the first step; the following section describes the second step.

\begin{figure*}
    \centering
    \includegraphics[width=\textwidth]{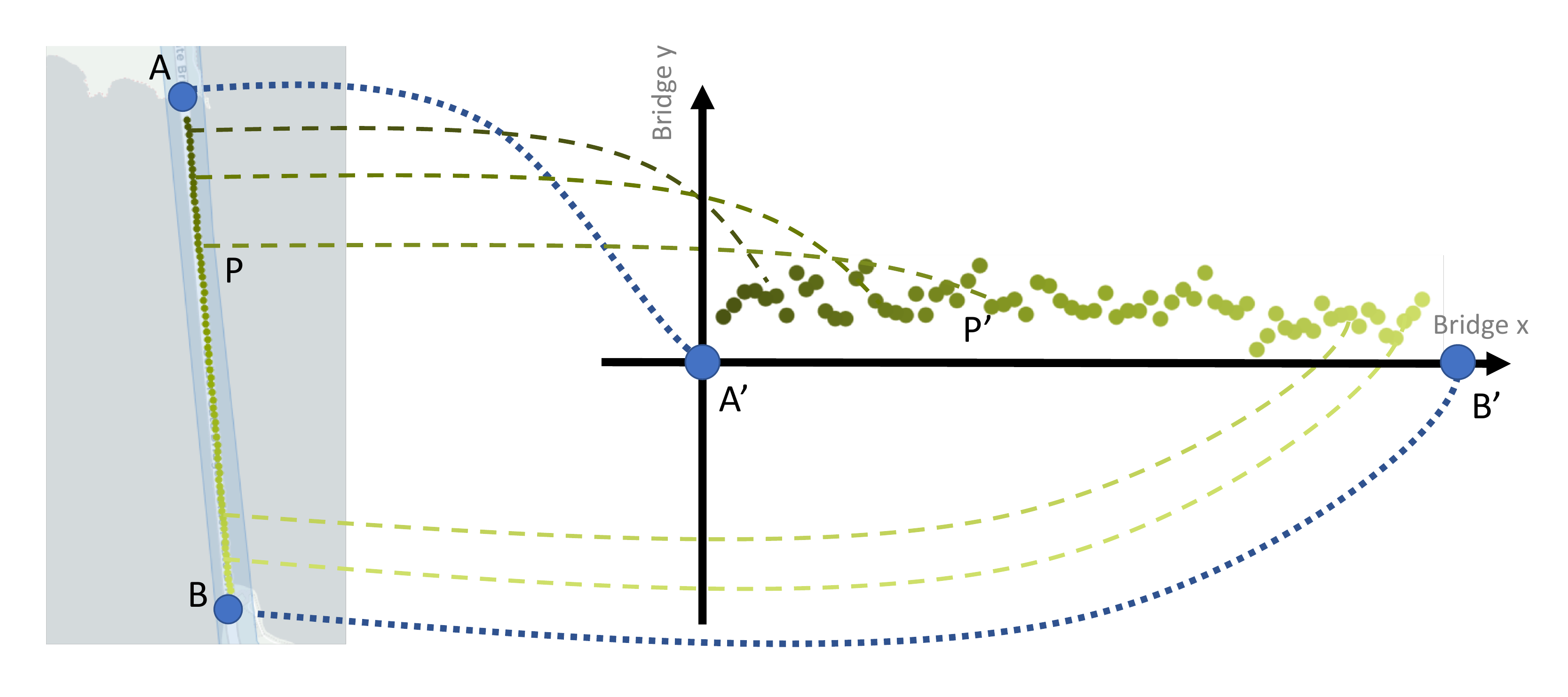}
    \caption{Illustration of transforming GPS points to bridge coordinates. The vehicle (mobile sensor) collects samples as it travels on the bridge from lat-long points $A$ to $B$ (left panel). The gradient of the points indicates the temporal dimension: the point with the lightest color is the most recent. On the right panel, the GPS points are mapped from lat-long values to a bridge coordinate system where $A'$ and $B'$ represent points at each end of the bridge and correspond to GPS points $A$ and $B$, respectively. Note that in the real application, the location of the towers were used as reference points $A$ and $B$, since our analysis only considers the main span of the bridge that has a total length of $1280\,\mathrm{m}$.}
    \label{fig:real_gps}
\end{figure*}

\begin{figure*}
    \centering
    \includegraphics[width=0.8\textwidth]{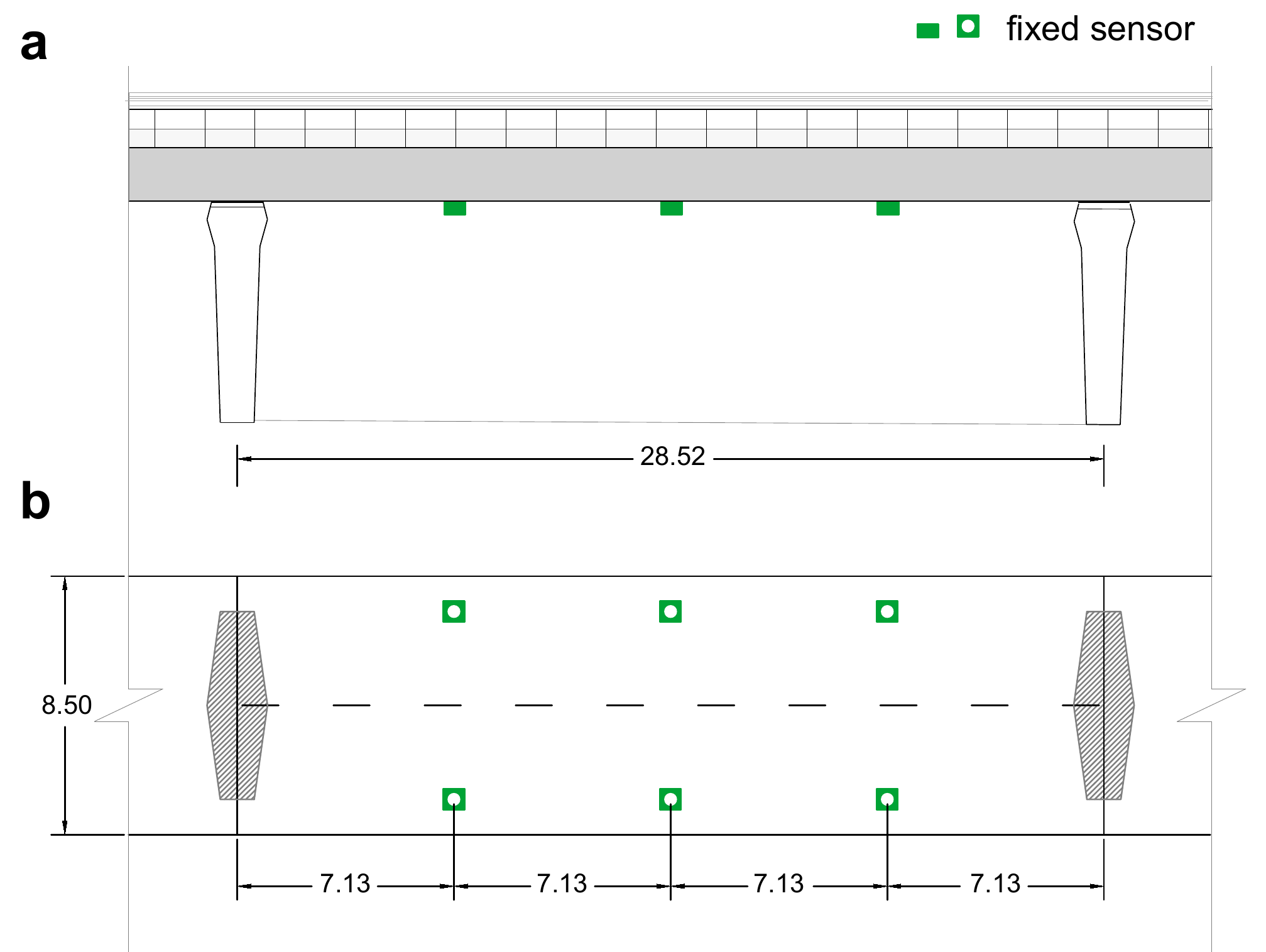}
    \caption{Arrangement of the network of fixed accelerometers installed on the short highway bridge in Ciampino, Italy. All dimensions in meters.}
    \label{fig:Ciampino_fixed}
\end{figure*}

Fig.~\ref{fig:real_gps} illustrates the first step of the process in which the GPS data are mapped to points on the bridge. For this, it is helpful to use a digital map to select points at the corners of roadway of the bridge which act as a frame of reference; these points should coincide with the start and end of the bridge. With these reference points established, the haversine formula can be used to calculate distances between GPS points and transform them into another coordinate system. The bridge coordinate system is viewed as a 2D Cartesian space defined based on the reference points. Fig.~\ref{fig:real_gps} shows how the reference points $A$ and $B$ on a map (satellite view) are mapped to points $A'$ and $B'$ in the bridge coordinate system. When possible, it is important to verify that the calculated distances, e.g., $\overline{A'B'}$, are consistent with the known geometry of the bridge. In addition, the quality of the GPS data should be inspected. Once the sensor positions are represented in the bridge coordinate system, the signal-to-noise ratio (SNR) of the points will vary for each dimension; generally, for smartphones, the SNR will be significantly higher in the primary traveling direction (parallel to the roadway, ``Bridge $x$'' in Fig.~\ref{fig:real_gps}). In this introductory application, it is sufficient to use only one dimension, e.g., $x$-coordinates, when describing the locations on the bridge at which data is being collected. In further applications, it may be necessary to use both dimensions.

\subsection*{Spatial Segmentation of a Bridge}
A fixed number of discrete, uniform overlapping segments are defined over the length of the bridge. The start and end of the spatial segmentation scheme are consistent with the reference GPS points chosen earlier to coincide with the bridge coordinate system. Fig.~\ref{fig:SpatialGrouping} shows discrete segments along the length of the bridge which have two defining parameters: the segment length, $c$,  and the distance between the centers of adjacent segments, $\Delta s$. The overlap of two adjacent segments is then a function of $c$ and $\Delta s$, namely, $c_o = c - \Delta s$.. The bridge segments, $\bar{s}= [s_1, s_2,...,s_M ]$, collectively define an interval in one-dimensional space, $[s_1-\frac{c}{2},s_M+\frac{c}{2}]$. Connecting these to the measurements, for the $i$th bridge crossing, there is a discrete set of $r_{ij}$ points that fall inside each bridge segment.

When selecting these parameters, it is important to consider the vehicle speeds, sensor sampling rates, and the geometry of the bridge, among other factors.
The spatial segmentation of the Golden Gate Bridge was set to have a large number of wide segments. A large number of segments (larger $M$ and smaller $\Delta s$) increases the number of total entries in the frequency candidates vector, $\hat{f}_s$ (see below), which improves the resolution of subsequent PDF estimates. Furthermore, wide segments (larger $c$) improve overall robustness to noise; if the frequency net in space is too small, then noisy, spontaneous frequencies, that are unrelated to the bridge's vibrations, may appear to be statistically significant. For the primary application (controlled data), the spatial segmentation scheme was set to have segment centers separated by $10$ meters, or $129$ equally-spaced segments ($\Delta s = L/129$) and the width of each segment was set to $258$ meters ($c = L/5$, $20\%$). For the ridesourcing data (uncontrolled), the width of each segment was $c = 3L/20$ ($15\%$). See the SI for additional analyses testing the sensitivity of the methods with regards to $\Delta s$ and $c$.

\subsection*{Determination of the Most Probable Modal Frequencies (MPMFs)}

\begin{table*}[t]

\centering
\fbox{\begin{minipage}{0.8\paperwidth}
\begin{enumerate}
    \item
    Vertical acceleration signal $x_i (t)$ and linear position in bridge coordinates $r_i(t)$ is extracted from the measurements for $i = 1, 2 \dots N$ datasets.
    \item
    Acceleration signal is filtered and downsampled to the desired frequency $f_\mathrm{cut}$. Time and space are discretized accordingly: $t_{ij} = j / f_\mathrm{cut}$ and $r_{ij} = r_i (t_{ij})$ with $j = 1, 2 \dots M_i$. Note that time and space discretization along with the value of $M_j$ can differ among the measurements in the dataset if vehicles travel at different speed. Note that this is not an issue because a common spatial segmentation of the bridge is implemented later. The Golden Gate Bridge analyses used $f_\mathrm{cut} = 0.5\,\mathrm{Hz}$.
    \item
    Synchrosqueezed wavelet transform is calculated for each measurement separately, resulting in the $T_{x_i} (f, t_{ij})$ time-frequency representation of the signals. Morlet wavelets were employed in this study.
    \item 
	The frequency-time pairs are mapped to frequency-space pairs, i.e., $(f,t_{ij})\rightarrow{}(f,r_{ij})$. This change of variables remaps the synchrosqueezed wavelet transform accordingly $T_{x_i} (f,t_{ij})\rightarrow{}T_{x_i} (f,r_{ij})$.
    \item 
	Define  $P_i (f,r_{ij})$ as the local maxima of  $|T_{x_i} (f,r_{ij})|$  $\forall i,j$
    \item 
	Select a value for $\alpha$. Compute $F_{P_i} (f|r_{ij})$, the empirical CDF of $P_{x_i} (f,r_{ij})$ $\forall i,j$. Define the statistical ridges as the values in the upper $\alpha$-percentile: $P_i (f_{1 - \alpha},r_{ij} )$ where $f_{1 - \alpha}$ are the values associated with $F_{P_i} (f_{1 - \alpha}| r_{ij} )>1-\alpha$
    \item 
	Aggregate the local frequency maxima using the summation $P_N (f,s_m)=\sum_{i=1}^{N} \sum_{r_{ij} \in [s_m-\frac{c}{2},s_m+\frac{c}{2})} P_{x_i}(f_{1 - \alpha},r_{ij})$
    \item
	Select a value for $N_R$. Define $\hat{f}_{s_m} $ as the extracted frequency vector which contains the frequencies corresponding to the $N_R$ largest values in $P_N (f,s_m )$. Combine all $M$ vectors to construct the frequency candidate vector $\hat{f}_s = [\hat{f}_{s_1}, \hat{f}_{s_2},..., \hat{f}_{s_M}] $ (this constitutes aggregation in space).
	\item
	Finally, estimate the PDF of $\hat{f}_s$ using kernel density estimation, from which the most probable modal frequencies (MPMFs) are determined. For reference, in this study Gaussian kernels were used with a bandwidth equal to 1\% of the frequency range.
\end{enumerate}
\end{minipage}}
\caption{Summary of the methodology for computing MPMFs for $N$ datasets.}
\label{tab:methods}

\end{table*}

The space-frequency representation of the signal ($T_{x_i} (f,r)$) is used as a basis of a statistical aggregation method that aims to select frequencies that are present consistently in the bridge vibrations and among the vehicle trips in our experiments. Note that a later extension of our method would easily allow to focus on frequencies that are present only in some locations, following the pattern of the associated spatial mode. The end result of our method is a probability distribution with peaks representing possible modal frequencies, from which the most probable modal frequencies (MPMFs) are identifiable.

The first step is to identify the local maxima within $|T_{x_i}(f,r)|$, which are related to the ridges of the synchrosqueezed wavelet transform (see the followig section for further details on this process).

Start with a set containing all local maxima of $|T_{x_i}|$ for each discrete location $r_{ij}$ for each bridge crossing. Only statistically significant peaks from each set will be kept, i.e., those with a prominence in the top $\alpha$ percentile of the corresponding empirical cumulative distribution function (CDF). The result is a family of piece-wise ridges for each dataset, as illustrated in Fig.~\ref{fig:wsst_max}a. The process is repeated for all $N$ datasets. Then the ridges inside each bridge segment, $s_m \in \bar{s}$ are aggregated for each bridge crossing.

Next, with a shared bridge segmentation scheme, the prominence values of all statistically significant ridges are aggregated among the whole dataset as well. In other words, all datasets are aggregated (see Fig.~\ref{fig:wsst_max}) to produce a matrix of the cumulative prominence values, $P_N$, with respect to all frequencies, $f$, and all segments, $\bar{s}$. Then, for each $s_m\in \bar{s}$, the prominence values are sorted in descending order, and the frequencies associated with the top $N_R$ values are stored in a vector $\hat{f}_{s_m}$, which is assigned to $s_m$. These vectors are combined to create the frequency candidate vector $\hat{f}_s$, which has $M \times N_R$ entries. 

The final probability distribution function (PDF) (shown in Fig.~\ref{fig:wsst_max}c as a histogram) is produced using the frequency candidate vector. This empirical PDF is expected to be multimodal, with each mode representing a possible vibrational frequency. The MPMFs can be defined as a subset of the possible modal frequencies having the largest amplitudes. At this stage, the problem of determining the MPMFs, relies on the method used for estimating the PDF, as the MPMFs are local maxima in the PDF. A kernel density estimation is used to fit a smooth, nearly continuous PDF; this improves the precision with which MPMFs are estimated.

The procedure for computing the MPMFs for $N$ datasets is summarized in Table~\ref{tab:methods}. In addition to defining bridge segmentation parameters, it is necessary to select parameters $\alpha$ and $N_R$ which impact ridge estimation and spatial frequency aggregation. In practice,  $\alpha \leq 0.05$ and $N_R = 5$ were used.

\subsection*{Piecewise Wavelet Ridges}

A ridge is a sequence of frequencies, $f(r)$, that follow a curve and trace maximum amplitudes ~\cite{delprat1992asymptotic, mallat1992characterization, guillemain1996characterization,carmona1997characterization}. Ridge extraction methods usually consider either the transform's modulus or its instantaneous phase. In applications with noisy data, modulus-based methods are often preferred \cite{staszewski1997identification, zhong2005phase}. Ridge extraction methods generally look for a smooth function (or set of functions), $r\rightarrow{}|T_x(f_R(r),r)|$, that concentrates most of the energy in the time-frequency domain. The nature of the estimated ridges rely on the complexities of the underlying signal components as well as the extraction method. For example, if a signal has only one component, $x_1(t) = A_1(t)cos(2\pi\varphi_1(t))$, its true ridge is unique and equivalent to the instantaneous frequency function, $f_R(t) = \varphi_1'(t)$ \cite{jiang2017instantaneous}.

In these applications, the signal has numerous underlying components, which have distinct characteristics that influence ridge extraction: (i) the components have time-invariant frequencies (a direct result of  Eq.~(2) in the SI); (ii) the components' amplitudes are intermittent; they are, by nature, stochastic and can vary quickly in time as they rely on an unknown dynamic excitation ($p(t)$ of  Eq.~(1) in the SI); (iii) as a corollary, some components are expected to be present in one data set, but absent in another. These characteristics call for a multiridge detection technique \cite{carmona1999multiridge,wang2013matching} that is capable of rapid ``switching'' between components whenever one disappears or another emerges. The approach used here is motivated by the method developed by \cite{mcaulay1986speech} for speech analysis, which can accommodate rapid changes in spectral peaks, i.e., sudden ``births'' and ``deaths'' of components in time.

\subsection*{Robustness analysis}

To estimate the robustness of the methods, all analyses were repeated for random samples selected from the data. For $N_S = 1, 5, 10, 20, 30, 40, 50, 60, 70, 80$ and $90$, $N_S$ bridge crossings were selected at random and analyses were repeated considering this random sample as an input. For each $N_S$ value, random sampling and estimation is repeated $100$ times; in each case, the top frequency is selected (mode of the histogram created from the corresponding $\hat{f}_s$ values) and compared with the candidate real vibrational frequencies of the Golden Gate Bridge. The average error is calculated based on the closest candidate and also count the number of times the error is less then 5\%.

Furthermore, the bridge-vehicle system is simulated and the signal is analyzed with the same methodology. This process yields similar results to what was observed in the real data, as shown in the Supplementary Information (Figs.~S1 and~S2).

\subsection*{Data Availability}

The "controlled data" that support the findings of this study are available from the corresponding author upon reasonable request.

The "uncontrolled ridesourcing data" that support the findings of this study are available from UBER but restrictions apply to the availability of these data, which were used under license for the current study, and so are not publicly available. Data are however available from the authors upon reasonable request and with permission of UBER.

The "partially-controlled data" that support the findings of this study are available from ANAS Sp.A. but restrictions apply to the availability of these data, which were used under license for the current study, and so are not publicly available. Data are however available from the authors upon reasonable request and with permission of ANAS Sp.A.

\subsection*{Code Availability}

The code to replicate this research can be requested from the corresponding author.

\newpage
\normalsize

\twocolumn[  
    \begin{@twocolumnfalse}
    
		\begin{center}
			{\huge \bf Supplementary Material \\[2.5ex]}
		\end{center}
		
    \end{@twocolumnfalse}
]

\renewcommand{\thefigure}{S\arabic{figure}}
\renewcommand{\thetable}{S\arabic{table}}

\setcounter{figure}{0}
\setcounter{table}{0}

\section*{Brief Review of Underlying Structural Dynamics}

Bridges are structural systems that respond dynamically to excitation forces, e.g., traffic, wind, etc. in accordance with the wave equation. In particular, the equation of motion defines this relationship explicitly in terms of the structure's physical properties

\begin{equation}
  \sum_{j=1}^N m_{ij} \ddot u_j(t) + \sum_{j=1}^N c_{ij} \dot u_j(t) + \sum_{j=1}^N k_{ij} u_j(t) = p_i(t)
  \label{eq:dynamics}
\end{equation}

where $\mathbf{m}$, $\mathbf{c}$, and $\mathbf{k}$ are the discrete-space mass, damping, and stiffness matrices and $\ddot{u}_i(t)$, $\dot{u}_i(t)$, $u_i(t)$ (with $i = 1,2\dots N$) are the accelerations, velocities, and displacements of the structure at $N$ specified degrees of freedom, and the dynamic forces are given by $p_i(t)$. In summary, $\mathbf{m}$, $\mathbf{c}$, and $\mathbf{k}$ are $N\times N$ matrices and $u(t)$, $\dot{u}(t)$, $\ddot{u}(t)$, and $p(t)$ are $N\times 1$ vectors. The current work focuses on \emph{vertical} displacements and the components are indexed accordingly using subscript $i$, i.e.~$u_i (t)$ describes the vertical displacement of the bridge at location $r \equiv \delta i$ (where $\delta$ is the spatial discretization used in the model). In this study, a very large number of degrees of freedom, $N$, is considered to attain a very fine discretization in space, as this is consistent with the measurements of a mobile sensor network.

For linear structural systems\footnote{In practice, it is difficult to confirm linear structural behavior because the internal and external dynamic forces are unmeasured and unknown. Suspension and stay-cable bridges, in particular, can exhibit strong geometric nonlinearities, especially in tower and cable dynamics. Regular bridge traffic is considered an ``operational load'' or an ``ambient excitation'' in the structural health monitoring community. Under ``ambient vibration'' conditions, i.e., as long as the excitation amplitudes remain relatively low, the bridge displacements are not large and the measured responses on the bridge deck are well characterized by the linear equation of motion \mbox{\cite{abdel1976dynamic,abdel1978ambient,abdel1985ambient,bendat1980engineering, bendat2011random}}. This approach is used widely and is embedded within numerous benchmark approaches such as \mbox{\emph{ERA-NExT}} \mbox{\cite{james1993natural}} and \mbox{\emph{FDD}} \mbox{\cite{ brincker2001modal}}.}, the dynamic response can be represented as a summation of vibration modes \cite{chopra2001dynamics} that are the solution of the homogeneous equation (i.e. the $p(t) = 0$ case). These can be written as $\Phi_{d} q_{d} (t)$ for $d = 1,2,\dots N$, where $q_{d} (t) = \Re e^{i \Omega_{d} t}$ are scalar harmonic functions, and $\Phi_{d}$ are $N\times 1$ vectors obtained as a solution of the generalized eigenvalue equation $-\Omega_{d}^{2} \mathbf{m} \Phi_{d} + i \Omega_{d} \mathbf{c} \Phi_{d} + \mathbf{k} \Phi_{d} = 0$. The eigendecomposition results correspond to the modal properties of the structure: mode shapes are given by the $\Phi_{d}$ vectors, while the eigenvalues $\Omega_{d} \equiv \omega_{d} + i \zeta_{d}$ give modal frequencies $\omega_{d}$ and damping rates $\zeta_d$. The solutions of Eq.~(\ref{eq:dynamics}) can then be represented as a linear combination of individual vibrational modes:

\begin{equation}
  u_j(t) = \sum\limits^N_{d=1}c_{d} (t) \Phi_{dj} q_d(t)
  \label{eq:modes}
\end{equation}

For a system described by $N$ degrees of freedom, there can be up to $N$ vibrational modes contributing to its movement at any time\footnote{Theoretically, there will always be $N$ terms present in Eq.~(\ref{eq:modes}). For efficient computation, modal truncation is often practiced, in which the summation includes only the first $\bar{N}$ terms where $\bar{N} \ll N$ and serves as an accurate approximation for $u(t)$. In other words, modal truncation assumes that the first $\bar{N}$ modes dominate the response.}. In the homogeneous case, the $c_{d}$ amplitudes are constants determined by the initial conditions, while in the general case with time-varying $p(t)$ excitation, the $c_{d}(t)$ coefficients will be time dependent as well. Nevertheless, in many cases, the rate of change of the $c_{d} (t)$ amplitudes is slow compared to the time-frame determined by the corresponding $\omega_{d}$ modal frequency, thus treating the behavior of the bridge as a sum of vibrational modes with time-varying amplitudes is justified, with each mode having a different contribution to the response at different times and locations.
The spatial component is often well modeled as a sinusoid as well, i.e.~the $j$th component in $\Phi_d$ is given by $\Phi_{dj} = \sin{2\pi \delta j / \lambda_d + \varphi_d}$, where $\lambda_d$ and $\varphi_d$ are the wavelength and phase of mode $d$.

Note that this representation is consistent with that given in Eq.~(1) in the main text. We gain the form in Eq.~(1) by switching to the continuous variable $r$ to represent space, adding the noise term $e(t)$ and incorporating damping ($\zeta_d$) and the effect of continuous excitations from $p_i (t)$ in the combined amplitude term $A_d (t)$.
The presence of a modal response in $u_j(t)$ is amplified when (i) the dynamic force has a spatial distribution that is strongly correlated to the mode shape; and/or (ii) the dynamic force has spectral content that is very close to a modal frequency.

In summary, the physical response, e.g., $u_j(t)$, is two-dimensional: it depends on space and time. Modal analysis shows that the response to dynamic loads has a unique spatial-spectral composition based on intrinsic modal vibration properties. The underlying vibration modes can be observed by analyzing measurements of the physical response, i.e., sensor data.

\section*{Wavelet-Based Time-Frequency Representation of Recorded Signal}

A primary objective of the analysis is to accurately determine the frequency content within the measured vehicle-bridge response. The structural response of the bridge is sampled in time and space through mobile sensing. The resulting signal may be non-stationary and non-linear, with frequency content that is highly dependent on time. Therefore, to avoid improper assumptions on the physical and statistical nature of the signal, a time-frequency representation (TFR) is needed to evaluate the underlying frequency components. A plethora of TFR methods have been employed for various SHM applications such as the Short-time Fourier Transform \cite{rioul1991wavelets}, Empirical Modal Decomposition \cite{huang1998empirical,flandrin2004empirical}, Wigner-Ville distribution \cite{claasen1980wigner}, and the Wavelet Transform \cite{daubechies1996nonlinear}

Synchrosqueezing is an algorithm applicable to TFRs of signals with time-varying spectral characteristics. The technique was introduced for wavelet transforms by \cite{daubechies2011synchrosqueezed} and further mathematical details and applications have been discussed in \cite{thakur2013synchrosqueezing,jiang2017instantaneous}. In this framework, the time-series signal $x(t)$ is represented in the general form 

\begin{equation}
  x(t) = \sum\limits^D_{d=1}x_d(t)+e(t)
  \label{eq:wsignal}
\end{equation}

where each signal component $x_d(t) = A_d(t) cos(2\pi\varphi_d(t))$ is a ``Fourier-like oscillatory mode'' with a time-dependent amplitude $A_d(t)$, time-dependent frequency $\dot{\varphi}_d(t)$, and signal noise $e(t)$ \cite{thakur2013synchrosqueezing,daubechies2011synchrosqueezed}. The ultimate goal is to obtain the amplitude
$A_d(t)$ of instantaneous frequency $\dot{\varphi}_d(t)$ for each $d$ . It is important to note the definition of $x(t)$ in Eqn.~(\ref{eq:wsignal}) is compatible with the modal analysis of $u_j(t)$ in Eqn.~(\ref{eq:modes}) and the $x_i(t)$ form presented as Eqn.~(1) in the main text. Most importantly, we assume that $\varphi_d(t) \equiv \omega_d t$ and that $A_d(t)$ is determined by the $\Phi_{dj}$ mode shape at the $r(t)$ location the vehicle is at time $t$.


Identifying the signal components then is achieved using a three-step process.

First, the TFR of the signal $x(t)$ is produced using the continuous wavelet transform (CWT)

\begin{equation}
  W_x(a,b) = \frac{1}{a}\int_{-\infty}^{\infty}x(t)\overline{\psi\left(\frac{t-b}{a}\right)}dt
\end{equation}

where $\overline{\psi (\xi)}$ is the complex conjugate of a selected mother wavelet, e.g., Morlet, $a$ is the scale parameter ($a>0$), and $b$ is the time offset parameter. Discretization of the time scale, $b$, follows that of the original discrete-time variable of the signal, $t$: a series of $K$ values separated by $\Delta t$.

Second, in the pursuit of an instantaneous frequency estimate, the time derivative of the CWT is used to produce the phase transformation $\omega_x(a,b)$ 

\begin{equation}
  \omega_x(a,b) = \frac{-i}{W_x(a,b)}\frac{\partial W_x(a,b)}{\partial b}
\end{equation}

where $i^2=-1$. The phase transform can provide the exact instantaneous frequency in the case where $\varphi_d(t)$ is constant. Overall, $\omega_x(a,b)$ is comparable to a ``FM demodulated frequency'' estimate at $(a,b)$\footnote{The partial derivative with respect to $b$, $\partial_bW_x(a,b)$, is equivalent to the one with respect to $t$, $\partial_tW_x(a,b)$  \cite{thakur2013synchrosqueezing}.}. Note that $\omega_x$ is undefined when $|W_x|= 0$; in practice, because of noise and other artifacts, a threshold $\gamma$ is used to mitigate instabilities by ignoring points where $|W_x| \leq \gamma$ \cite{thakur2013synchrosqueezing}.

Third, the time-scale plane is transformed to a time-frequency plane in a process called synchrosqueezing; in other words, the scale variable is reassigned to a frequency variable. This results in the Wavelet Synchrosqueezed transform of $x(t)$, obtained as:

\begin{equation}
T_x (f,b) = \int_{\{a:|W_x(a,b)|>\gamma\}}{} W_x(a,b)\delta(\omega_x(a,b) - f)a^{-1}da
\label{eq:wsstfinal}
\end{equation}

where $f$ is a frequency variable and $\delta()$ is the Dirac delta function. At this stage, $a,b$ and $f$ are usually discretized for efficient digital computation and equation (\ref{eq:wsstfinal}) is calculated as discrete summations. The discrete frequencies are distributed logarithmically, defined as $f_l = \frac{1}{K \Delta t} 2^{l \Delta f}$ where $\Delta f = \frac{1}{n_a-1}\log_2(K/2)$ for a signal with $K$ samples in time with $\Delta t$ sampling interval and $n_a$ desired discrete frequencies. For the sake of consistency among signals of different length, the frequency resolution is commonly defined by $n_{v}$, the number of ``voices'' in an octave, i.e.~in an interval in which the frequency is doubled: $n_{a} = L n_{v}+1$, where $L = \log_2 (M/2)$ is the number of frequency octaves possible in a signal of $M$ samples \cite{goupillaud1984cycle}.

\section*{Measurement Noise Sensitivity Analysis using Numerical Simulations}

Smartphones collect noisy IMU data. In order to provide guidance for the hyperparameter selection and establish sensitivity to measurement noise, we conducted a numerical simulation of the bridge-vehicle system based on identified modal characteristics of the Golden Gate Bridge from the latest comprehensive monitoring project \cite{pakzad2008design}. The bridge is modeled as a multi-degree-of-freedom linear system with similar geometries to the original structure. For operational, ambient vibrations, it is common practice to assume the structural system and material behave linearly. In order to take the vehicle-bridge interactions into account, a simplified modeling approach is adopted from \citet{eshkevari2020simplified}. In summary, the approach posits that due to the negligible weight of a sensing vehicle compared to a long-span bridge, the dynamic analysis of vehicle-bridge interacting system can be decoupled. This in turn, dramatically reduces the computational costs and model complexity while maintaining the accuracy of estimations high. \par

In this numerical study, different SNR levels are simulated in order to investigate the effect on the accuracy of estimations as well the minimum required number of trips to reach a predefined estimation confidence threshold. To capture the effect of the random traffic load, a random white spatio-temporal loading is applied on the bridge while the sensing vehicles are in motion \cite{eshkevari2020simplified}. Sensing vehicles are modeled as two degrees of freedom quarter-cars with linear mechanical properties, each randomly sampled from log-normal distributions with means set to the properties of a commercial vehicle. In addition, in accordance to the real sensing scenarios, vehicles adopt different constant speeds while sensing.

\begin{figure*}
    \centering
    \includegraphics[width=3in]{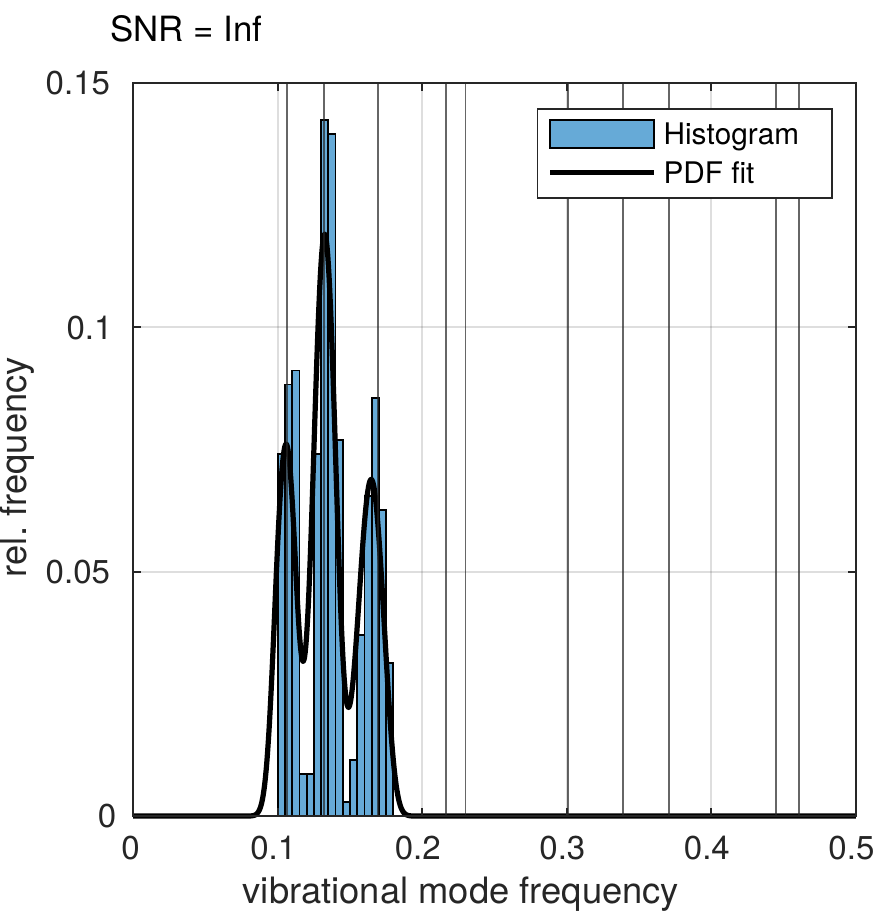} \quad
    \includegraphics[width=3in]{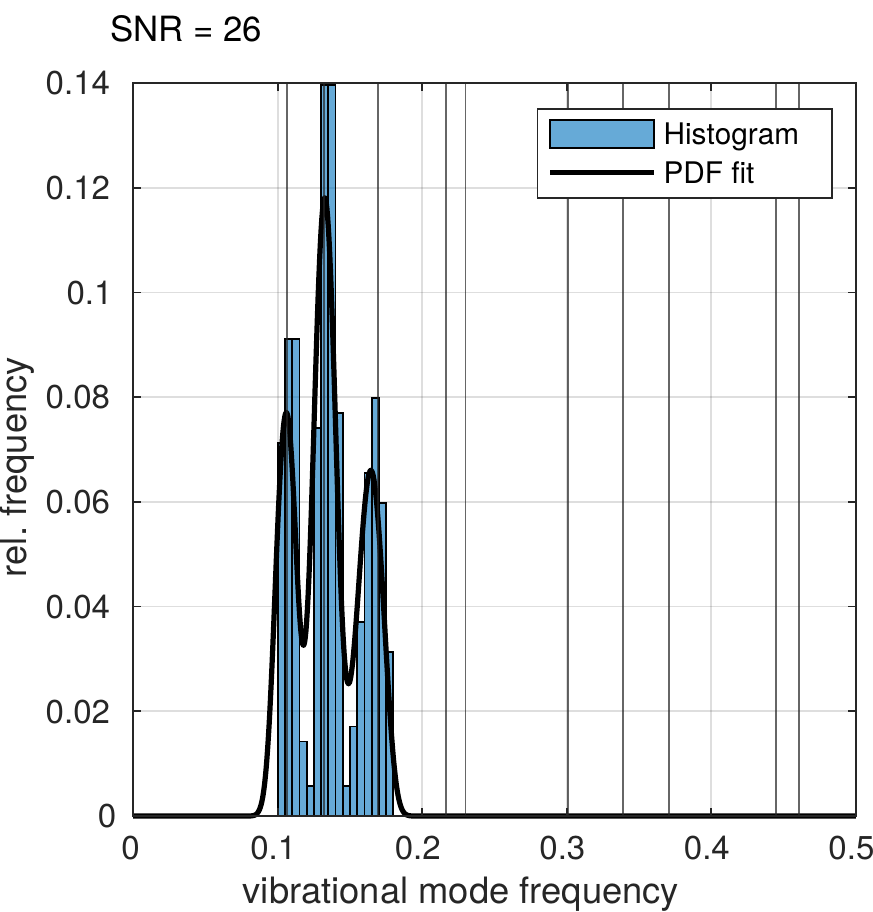} \\
    \includegraphics[width=3in]{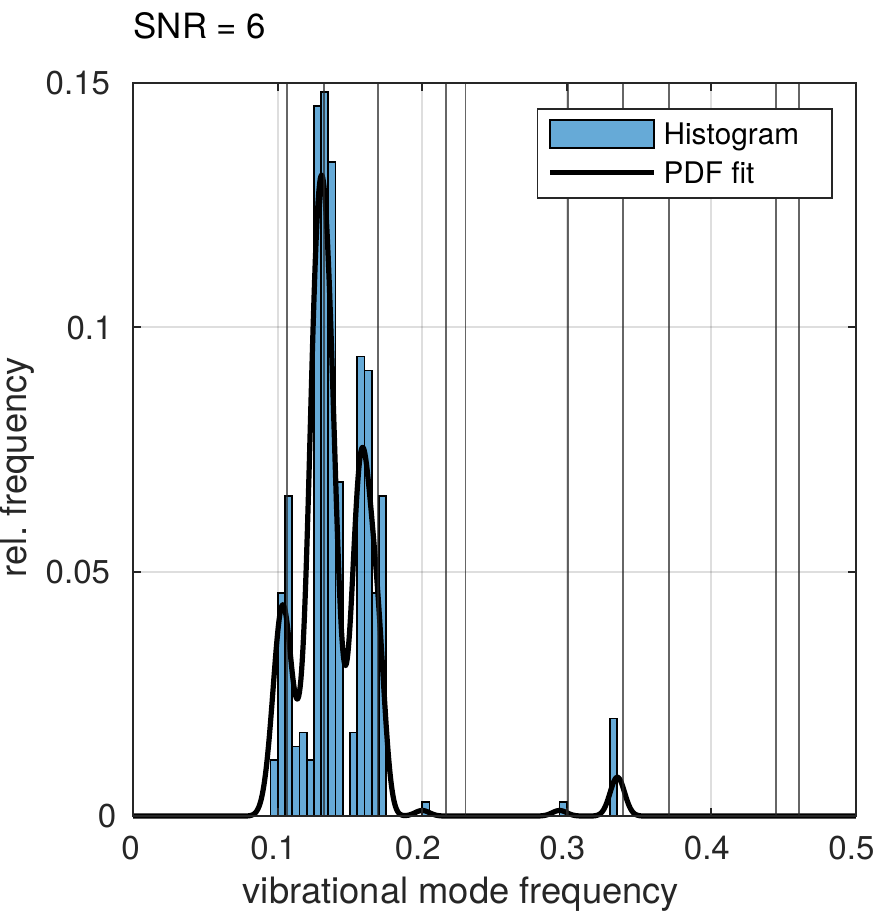} \quad
    \includegraphics[width=3in]{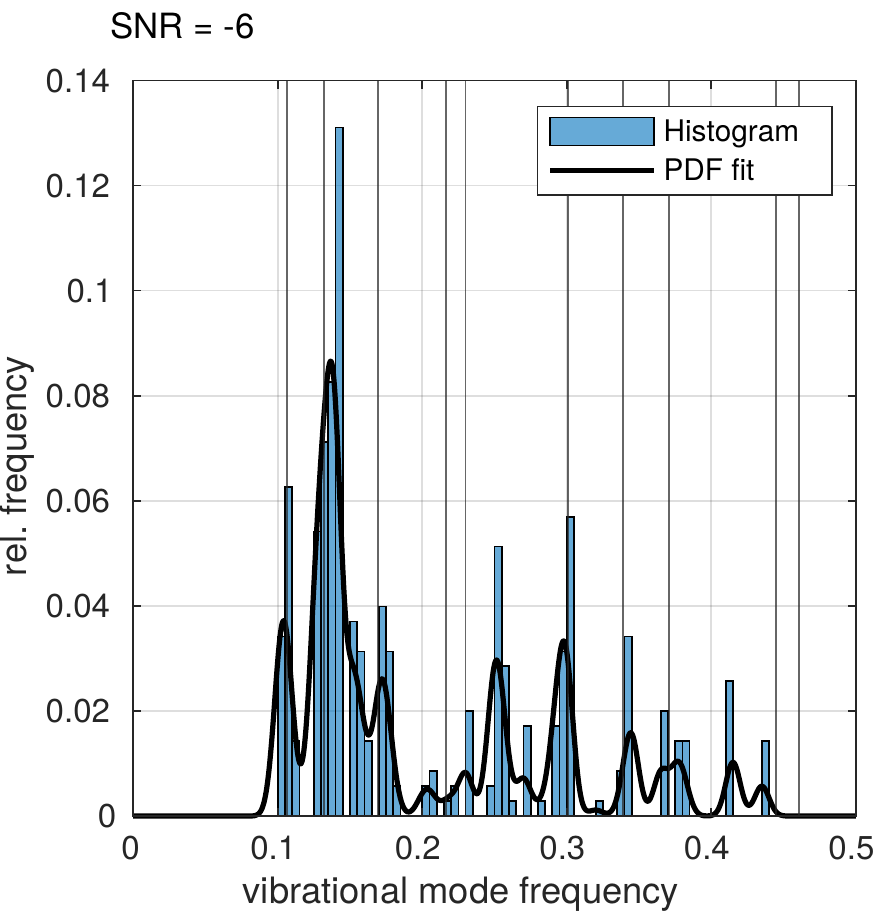}
    \caption{Histogram estimates of MPMFs with simulated data for varying SNR.}
    \label{sim_hist}
\end{figure*}

\begin{figure}
    \centering
    \includegraphics[width=3.2in]{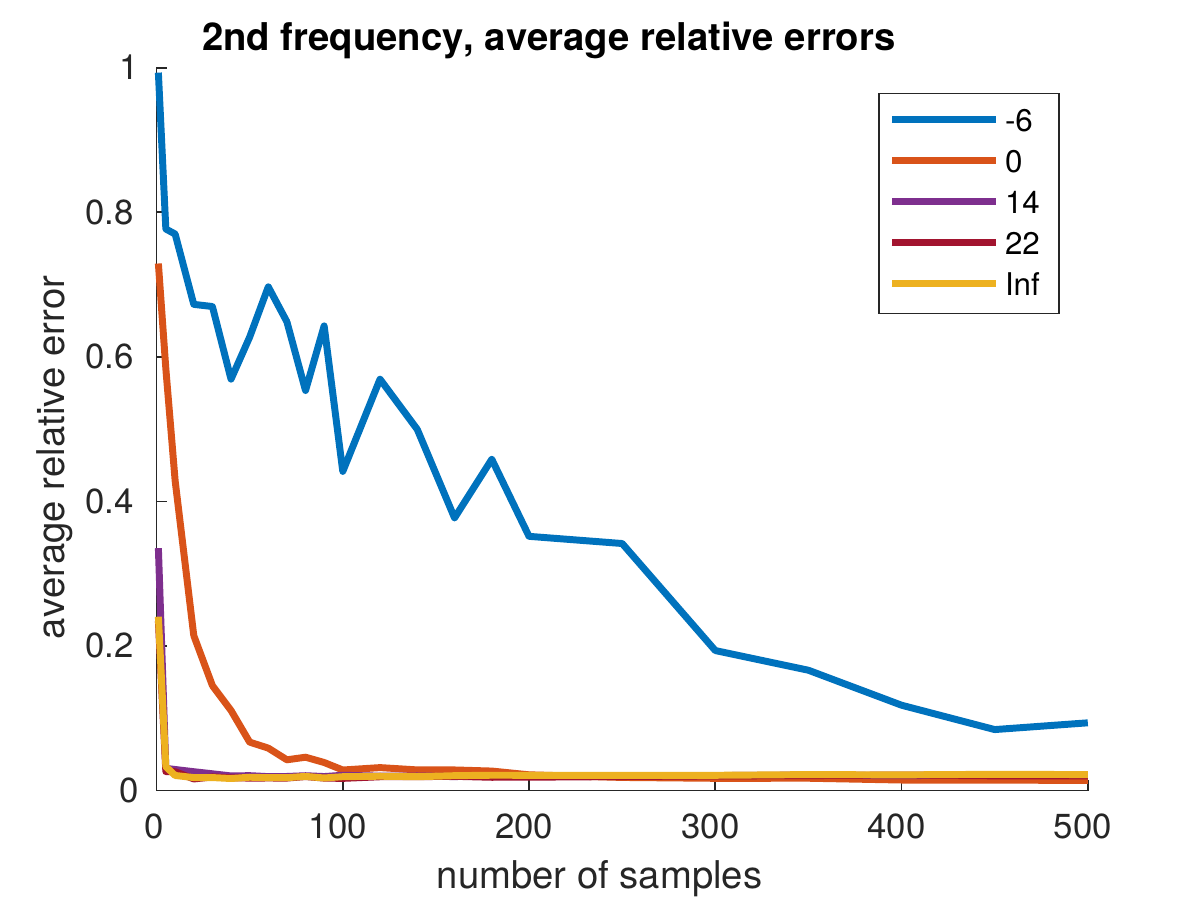}
    \caption{Relative error of estimation of the 2nd vibrational mode frequency as a function of the number of trip used with simulated data. Different colors represent varying SNR cases. In most cases, the estimation error decreases rapidly and then stays constant at a level that likely represents the limit of this estimation procedure. For especially high noise (SNR = 0 and -6), the decrease of errors is much slower, requiring a large number of bridge crossings to arrive at a reasonable estimate.}
    \label{sim_err2}
\end{figure}

As a result of this process, we arrive at similar estimates of MPMFs as for the main results displayed in Figs.~2 and~3 in the main text. We show these for four different cases of SNR in Fig.~\ref{sim_hist}, along with how errors in estimating the 2nd modal frequency decrease with increasing the number of bridge crossings considered in Fig.~\ref{sim_err2}.

\section*{Vehicle-Bridge-Road Interaction}
The dynamics of a vehicle crossing a bridge is characterized by the interaction of three systems: the vehicle suspension system, the bridge system, and the road surface. The influence of the vehicle-bridge-road interaction (VBRI) on measurements from mobile sensor networks has been studied widely  \mbox{\cite{cantieni1992dynamic,yang1995vehicle,yang1997vehicle, barella1995vehicle,yang2004hilbert,yang2004extracting,lin2005use,yang2009extracting,gonzalez2012identification,siringoringo2012estimating,yang2013frequency,eshkevari2020simplified}}. These studies have established the governing differential equations, constructed useful simplified models, provided informative simulations, and conducted real-world experiments, which have led to important discoveries. For example, the ``vehicle driving frequency'', which arises in certain vehicle scanning scenarios \mbox{\cite{yang1995vehicle,yang1997vehicle}}; and how low
vehicle speeds, stiff vehicle suspensions, and smooth road
surfaces can greatly simplify VBRI models \mbox{\cite{cantieni1992dynamic,barella1995vehicle}}.

Most commercially available vehicles have resonant frequencies within the range of $1-3$ Hz \mbox{\cite{olley1934independent,crolla1999olley}}, which may be close to modal frequencies of some short-span and medium-span bridges \mbox{\cite{yang1995vehicle,yang1997vehicle, barella1995vehicle,yang2004hilbert,yang2004extracting,lin2005use,yang2009extracting}}. In such cases, it may be desirable to remove these unwanted vehicle vibrations as they may interfere with the identification of bridge modal properties. For instance, if only one vehicle is used for vehicle scanning data, and one of its resonance frequencies coincides with a bridge modal frequency, it is more challenging to confirm bridge modal properties - it may be necessary to incorporate data collected by another vehicle. In general, source-separation methods, e.g., blind-source \mbox{\cite{cardoso1998blind,poncelet2007output}}, deconvolution \mbox{\cite{milliken2002chassis,mcgetrick2015experimental,SADEGHIESHKEVARI2020106733}}, etc., have demonstrated an ability to remove the contributions of the vehicle suspension system from the total vehicle response, such that the remaining signal corresponds to the bridge system.  The case of crowdsourcing data introduces unique stochastic elements to VBRI through the variety of vehicles and smartphones utilized. For example, when considering random vibrations caused by regular traffic on medium-span and long-span bridges, the effects of the dynamic forces induced on the bridge by the vehicle carrying the sensor are negligible \mbox{\cite{eshkevari2020bridge}}. In short, entropy is a desirable feature of crowdsourced vehicle-trip data which will help reduce bias in estimates of bridge properties. 

The studies in this paper did not explicitly consider VBRI, yet were still successful in identifying bridge modal frequencies. The main finding here is the persistent signal strength of bridge modal frequencies among the collected data sets. In the primary application on the Golden Gate Bridge, the frequencies of interest were known and below $0.5$ Hz. Therefore, downsampling and low-pass filtering eliminated vehicle dynamic effects. Free vibration tests conducted on the cars used in the controlled experiment confirmed fundamental vehicle frequencies around $1.7$ Hz. The same applied to the uncontrolled ridesourcing data. The application to the short-span highway bridge (partially-controlled data) presented a situation where the bridge modal frequencies overlapped with typical vehicle modal frequencies; in which case, it may have been desirable to account for unwanted VBRI effects. However, vehicle information was not available for this data set and there was insufficient ``off-bridge'' driving data to confirm individual vehicle modal frequencies. Nonetheless, the fundamental modal frequency of the bridge was accurately estimated with an error that decreased as the number of datasets increased.

Future work may leverage smartphone metadata on uncontrolled data factors such as smartphone mount, or vehicle  model, to help accurately characterize VBRI models. For instance, with sufficient data, acceleration measurements collected while a vehicle is \mbox{\emph{not}} on the bridge, can be helpful in identifying the vehicle dynamical system, e.g., transfer functions, and quantifying sensor measurement noise. There is also a need to study the synchronization problem that arises in the case of multiple moving sensors with independent sampling properties \mbox{\cite{marvasti2012nonuniform,zabini2016random,zabini2016inhomogeneous}}. Lastly, as vehicular networks emerge for urban sensing, e.g., vehicle-vehicle systems, the design of intercommunication systems should consider computation and energy costs related to data processing and transmission \mbox{\cite{taysi2012routing,rabiei2013rainfall,bazzi2015ieee,massaro2016car}}.

\section*{Impact on Bridge Service Life}
How will this new wealth of information impact the longevity of existing bridges? Crowd-sourced mobile sensors will provide a nearly continuous stream of information on structural modal properties, which is vital to condition assessments and damage identification frameworks for bridges \mbox{\cite{farrar1996damage,wahab1999damage,peeters2001one,yang2004extracting,nair2006time,yao2012autoregressive,cross2013long}}.  When the resulting information is incorporated into a bridge management plan, the benefits accumulate over the service life of a bridge.  This section investigates the potential of crowd-sourced monitoring approaches to increase bridge service life through a structural reliability analysis \mbox{\cite{frangopol1997life}}. 

A bridge owner's task is to manage degradation and allocated expenditures to ensure the bridge reaches its designed service life.  Here the bridge's condition (health) is measured by the reliability index $\beta$, which is constructed by integrating the states of individual structural components. By definition, the reliability index is inversely proportional to the failure probability of the bridge. Preventive maintenance activities increase $\beta$ with a factor that is proportional to the extent of the maintenance event. In a numerical setting, for a given bridge, $\beta$ (i.e., $\beta_0$) are initiated in accordance with its age and current condition (based on the most recent bridge monitoring results). Then a $\beta$ profile is generated over a period of time, depending on maintenance events and an inherent degradation rate. The result is called a \mbox{\textit{reliability profile}}.

In a generic model of an ageing bridge, the reliability index decays over time with a rate that depends on its construction quality, its environment, its exposure to the hazards,  among other factors \mbox{\cite{dissanayake2008reliability}}. The primary role of a structural health monitoring (SHM) system is to measure samples of the reliability profile and update the status of the bridge based on actual behavior. The ultimate goal of bridge maintenance is to maximize the point in time when the reliability index crosses threshold through applying preventive maintenance actions that result in a sharp increase in the reliability profile. Simultaneously, each maintenance action has an associated cost; therefore, from an asset management perspective it is important to balance these action costs with the bridge value.

The ideal bridge management system has complete information of structural conditions at any given moment in time for the entire service life of the bridge, i.e., a continuous or finely sampled reliability profile. In practice, it is prohibitively costly to achieve this volume and rate of information using modern SHM techniques. Crowdsensed data offer an opportunity to sample structural conditions at unprecedented rates and volumes, e.g., weekly, daily, hourly, etc., which can produce a highly discretized reliability profile. To quantify the value of this discretization, the reliability profiles of distinct classes of bridges are simulated. In reliability assessments, structural failure probabilities are typically derived with respect to a computational model that is updated based on actual SHM and environmental data \mbox{\cite{li2012reliability,baldomir2013reliability}}. More specifically, it is assumed that the crowdsourced data enables up-to-date knowledge on structural behavior using models that accurately reflect real-world observations. Note this assumption implies access to structural dynamics information beyond modal frequencies, which would result from long-term monitoring data, advanced SHM methods, AI, computational learning, etc., as mentioned in the \emph{Discussion} of the main text. 

Reliability profiles were generated for two bridge archetypes: (a) a typical bridge in the U.S., which is 43 years old (initial reliability index, $\beta_0$, randomly sampled from $U[5,6.4]$) and (b) a newly constructed bridge (initial $\beta_0$ randomly sampled from $U[8,9]$). For each bridge archetype, three management policies are considered: (i) no regular inspection or preventive intervention (PI), (ii) traditional approach in which PIs occur regularly in average every 15 year as planned and to a predefined extent (e.g., bearing replacements, pavement repairs, etc.), and (iii) crowd-sourcing approach in which a one-time PI is planned based on continuous SHM information of the bridge status and applied when the status reaches the service limit (a bridge is out of service when $\beta$ falls below $4.6$). 

For simplicity, the installation and maintenance costs of the sensor networks in policy (iii) are not considered. Note, the cost of monitoring in the crowd-sourcing framework (policy ii) is very low comparatively because the devices have already been purchased and the necessary mobility patterns are already in place. For an equal comparison, the maintenance parameters are adjusted such that the present value of the maintenance cost is identical between policies (ii) and (iii) (an $APR=6\%$ is considered). For each bridge archetype, the reliability modeling parameters such as initial reliability index $\beta_0$, degrading rate, and maintenance-caused improvement are randomly selected from distinct distributions as recommended in \mbox{\cite{frangopol2001reliability}} and $10,000$ reliability profiles are generated.

\begin{figure*}[h]
    \centering
    \includegraphics[width=6.4in]{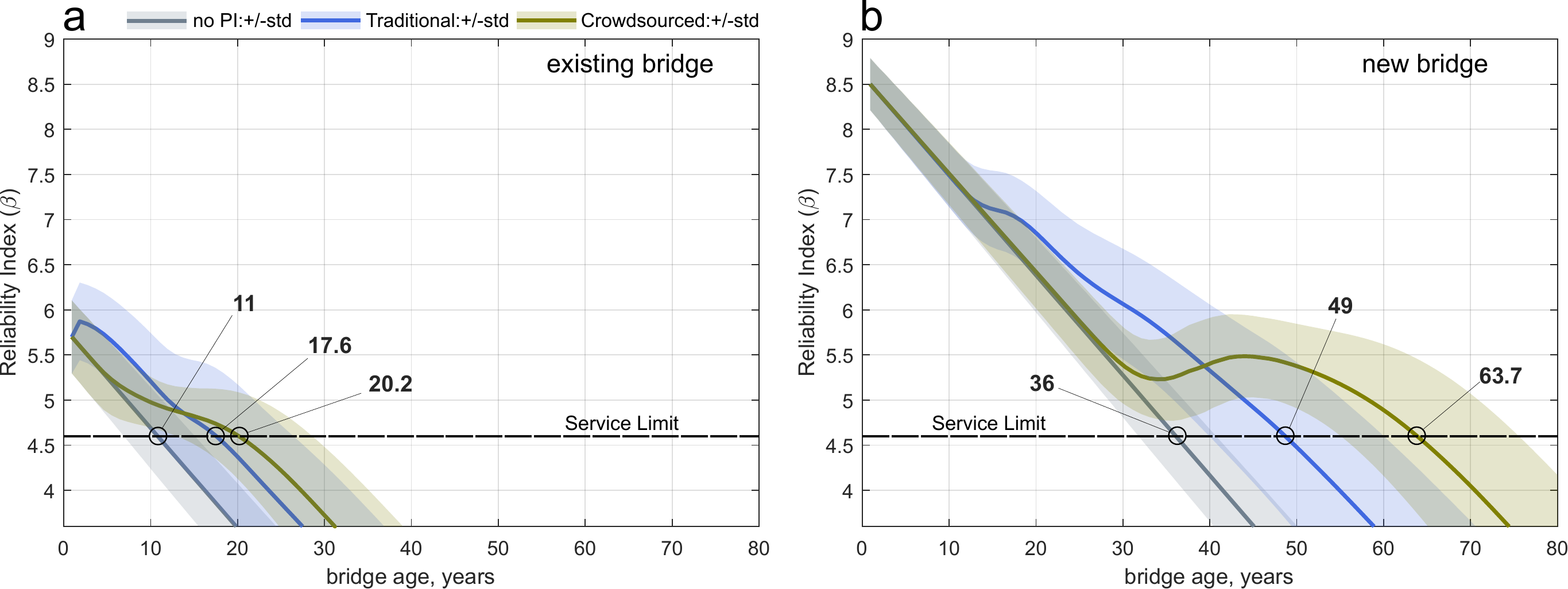}
    \caption{Reliability profiles of bridges based on Monte Carlo analyses: a) existing bridge at a typical age (43 years old); b) newly constructed bridge. Three policies are considered for each bridge archetype which have different effects on the bridge service life: (i) no preventive intervention (PI), (ii) traditional approach in which PI occurs in 15-year intervals, and (iii) crowd-sourcing approach in which a one-time PI is planned based on crowd-sourced information of the bridge status.  The bridge service life ends when the reliability index falls below the service limit ($\beta = 4.6$). The expected value for service life is highlighted by a circle marker with its corresponding value in bold. The crowdsourcing policy extends the service life of both bridge archetypes (regardless of the initial state). Note the life extension is significantly higher when the policy is adopted at the beginning of a bridge's operation:  14.5 year lifespan increase ($30\%$) for new bridges vs. 2.5 year lifespan increase  ($15\%$)  for typical bridges.}
    \label{MCS}
\end{figure*}

In Figure \mbox{\ref{MCS}}, the average reliability profiles under different policies are plotted plus or minus one standard deviation (shaded areas). The averaged profiles (solid lines) show the expected values for the reliability index. The intersection of the solid line and the service limit indicates the expected service life of the bridge. These values are highlighted in bold on each part of Figure \mbox{\ref{MCS}}. The results differ significantly between aged and newly constructed bridges regardless of the management plan. In both bridge archetypes, continuous monitoring enabled timely maintenance that extended the service life of a bridge compared to current methods without additional costs. On average, crowd-sourced data added 2.6 years of service to a typical bridge (a $15\%$ increase). The most substantial benefit was observed for a newly constructed bridge, whose service was extended by 14.7 years (a $30\%$ increase). 

These analyses are the first to quantify the cumulative benefits provided by frequent, large-scale bridge health monitoring. The continuous SHM information utilized to assess a one-time PI, would be based on long-term bridge health data including modal frequencies and other structural damage metrics. Modal frequencies are a gateway to determining additional modal properties (mode shapes, damping ratios, and higher modes) and structural features which have explicit relationships with certain types of bridge damage and deterioration, e.g., mode shape curvature, local stiffness, etc. \mbox{\cite{pandey1994damage,farrar1996damage,wahab1999damage,ruzzene1997natural,peeters2001one,gonzalez2012identification}}. Such  "damage-sensitive features" are important in the context of long-term structural behavior because large fluctuations, e.g., \mbox{$5-10 \%$}, of a bridge's modal frequency can be normal in some climates however do not necessarily represent a change in structural condition. Recent studies have successfully leveraged the environmental co-dependencies of modal frequencies to predict expected trends and detect unexpected ones based on as little as eight months of data \mbox{ \cite{liang2018frequency}}. For instance the changes in modal frequencies not explained by environmental data have been linked to bridge damage in real-world scenarios \mbox{ \cite{lee2005neural, sahin2003quantification,  ralbovsky2010frequency, kim2007vibration, fan2011vibration, peeters2001one, jin2016damage, liang2018frequency}}.

\small
\bibliographystyle{abbrv}
\bibliography{refs,refs_si}

\begin{thebibliography}{100}

\bibitem{abdel1976dynamic}
A.~M. Abdel-Ghaffar.
\newblock Dynamic analyses of suspension bridge structures.
\newblock Technical report, Earthquake Engineering Research Laboratory,
  California Institute of Technology, 1976.

\bibitem{abdel1978ambient}
A.~M. Abdel-Ghaffar and G.~W. Housner.
\newblock Ambient vibration tests of suspension bridge.
\newblock {\em Journal of the Engineering Mechanics Division}, 104(5):983--999,
  1978.

\bibitem{abdel1985ambient}
A.~M. Abdel-Ghaffar and R.~H. Scanlan.
\newblock Ambient vibration studies of golden gate bridge: I. suspended
  structure.
\newblock {\em Journal of Engineering Mechanics}, 111(4):463--482, 1985.

\bibitem{fhwa}
U.~F.~H. Administration.
\newblock National bridge inventory.

\bibitem{anjomshoaa2018city}
A.~Anjomshoaa, F.~Duarte, D.~Rennings, T.~J. Matarazzo, P.~deSouza, and
  C.~Ratti.
\newblock City scanner: Building and scheduling a mobile sensing platform for
  smart city services.
\newblock {\em IEEE Internet of Things Journal}, 5(6):4567--4579, 2018.

\bibitem{ARBTA2019}
A.~R. . T.~B. Association.
\newblock 2019 bridge report.
\newblock \url{https://tinyurl.com/y6fg636t}, August 2019.

\bibitem{baldomir2013reliability}
A.~Baldomir, I.~Kusano, S.~Hernandez, and J.~Jurado.
\newblock A reliability study for the messina bridge with respect to flutter
  phenomena considering uncertainties in experimental and numerical data.
\newblock {\em Computers \& Structures}, 128:91--100, 2013.

\bibitem{barella1995vehicle}
S.~Barella and R.~Cantieni.
\newblock Vehicle/bridge interaction for medium span bridges-research element 6
  of the oecd ir6 divine project.
\newblock In {\em Proc., 4th Int. Symp. Heavy Veh. Weights Dimensions Road
  Transp. Technol.}, pages 355--364, 1995.

\bibitem{bazzi2015ieee}
A.~Bazzi, B.~M. Masini, A.~Zanella, and G.~Pasolini.
\newblock Ieee 802.11 p for cellular offloading in vehicular sensor networks.
\newblock {\em Computer Communications}, 60:97--108, 2015.

\bibitem{bendat1980engineering}
J.~S. Bendat and A.~G. Piersol.
\newblock Engineering applications of correlation and spectral analysis.
\newblock {\em New York}, 1980.

\bibitem{bendat2011random}
J.~S. Bendat and A.~G. Piersol.
\newblock {\em Random data: analysis and measurement procedures}, volume 729.
\newblock John Wiley \& Sons, 2011.

\bibitem{brincker2007automated}
R.~Brincker, P.~Andersen, and N.-J. Jacobsen.
\newblock Automated frequency domain decomposition for operational modal
  analysis.
\newblock In {\em Conference proceedings: IMAC-XXIV: A conference \& exposition
  on structural dynamics}. Society for Experimental Mechanics, 2007.

\bibitem{brincker2001modal}
R.~Brincker, L.~Zhang, and P.~Andersen.
\newblock Modal identification of output-only systems using frequency domain
  decomposition.
\newblock {\em Smart materials and structures}, 10(3):441, 2001.

\bibitem{brock2018large}
A.~Brock, J.~Donahue, and K.~Simonyan.
\newblock Large scale gan training for high fidelity natural image synthesis.
\newblock {\em arXiv preprint arXiv:1809.11096}, 2018.

\bibitem{brunton2016discovering}
S.~L. Brunton, J.~L. Proctor, and J.~N. Kutz.
\newblock Discovering governing equations from data by sparse identification of
  nonlinear dynamical systems.
\newblock {\em Proceedings of the national academy of sciences},
  113(15):3932--3937, 2016.

\bibitem{cantieni1992dynamic}
R.~Cantieni.
\newblock Dynamic behavior of highway bridges under the passage of heavy
  vehicles.
\newblock 1992.

\bibitem{cardoso1998blind}
J.-F. Cardoso.
\newblock Blind signal separation: statistical principles.
\newblock {\em Proceedings of the IEEE}, 86(10):2009--2025, 1998.

\bibitem{carmona1997characterization}
R.~A. Carmona, W.~L. Hwang, and B.~Torr{\'e}sani.
\newblock Characterization of signals by the ridges of their wavelet
  transforms.
\newblock {\em IEEE transactions on signal processing}, 45(10):2586--2590,
  1997.

\bibitem{carmona1999multiridge}
R.~A. Carmona, W.~L. Hwang, and B.~Torr{\'e}sani.
\newblock Multiridge detection and time-frequency reconstruction.
\newblock {\em IEEE transactions on signal processing}, 47(2):480--492, 1999.

\bibitem{catbas2007limitations}
F.~N. Catbas, S.~K. Ciloglu, O.~Hasancebi, K.~Grimmelsman, and A.~E. Aktan.
\newblock Limitations in structural identification of large constructed
  structures.
\newblock {\em Journal of Structural Engineering}, 133(8):1051--1066, 2007.

\bibitem{pewsmartphone}
P.~R. Center.
\newblock Smartphone ownership is growing rapidly around the world, but not
  always equally.
\newblock
  https://www.pewresearch.org/global/2019/02/05/smartphone-ownership-is-growing-rapidly-around-the-world-but-not-always-equally/,
  2019.

\bibitem{chopra2001dynamics}
A.~K. Chopra.
\newblock {\em Dynamics of structures: theory and applications to earthquake
  engineering}.
\newblock Prentice-Hall, 2001.

\bibitem{claasen1980wigner}
T.~Claasen and W.~Mecklenbrauker.
\newblock The wigner distribution—a tool for time-frequency signal analysis.
\newblock {\em Philips J. Res}, 35(3):217--250, 1980.

\bibitem{crolla1999olley}
D.~Crolla and R.~King.
\newblock Olley's “flat ride” revisited.
\newblock {\em Vehicle System Dynamics}, 33(sup1):762--774, 1999.

\bibitem{cross2013long}
E.~Cross, K.~Koo, J.~Brownjohn, and K.~Worden.
\newblock Long-term monitoring and data analysis of the tamar bridge.
\newblock {\em Mechanical Systems and Signal Processing}, 35(1-2):16--34, 2013.

\bibitem{daubechies1996nonlinear}
I.~Daubechies.
\newblock A nonlinear squeezing of the continuous wavelet transform based on
  auditory nerve models.
\newblock {\em Wavelets in medicine and biology}, pages 527--546, 1996.

\bibitem{daubechies2011synchrosqueezed}
I.~Daubechies, J.~Lu, and H.-T. Wu.
\newblock Synchrosqueezed wavelet transforms: An empirical mode
  decomposition-like tool.
\newblock {\em Applied and computational harmonic analysis}, 30(2):243--261,
  2011.

\bibitem{delprat1992asymptotic}
N.~Delprat, B.~Escudi{\'e}, P.~Guillemain, R.~Kronland-Martinet,
  P.~Tchamitchian, and B.~Torresani.
\newblock Asymptotic wavelet and gabor analysis: Extraction of instantaneous
  frequencies.
\newblock {\em IEEE transactions on Information Theory}, 38(2):644--664, 1992.

\bibitem{dissanayake2008reliability}
P.~Dissanayake and P.~Karunananda.
\newblock Reliability index for structural health monitoring of aging bridges.
\newblock {\em Structural Health Monitoring}, 7(2):175--183, 2008.

\bibitem{elhattab2019extraction}
A.~Elhattab, N.~Uddin, and E.~OBrien.
\newblock Extraction of bridge fundamental frequencies utilizing a smartphone
  mems accelerometer.
\newblock {\em Sensors}, 19(14):3143, 2019.

\bibitem{en19912}
B.~EN.
\newblock 2: 2003, eurocode 1: Actions on structures-part 2: Traffic loads on
  bridges.
\newblock In {\em ICS}, volume~91, pages 93--040, 1991.

\bibitem{eshkevari2020bridge}
S.~S. Eshkevari, L.~Cronin, S.~N. Pakzad, and T.~J. Matarazzo.
\newblock Bridge structural health monitoring using asynchronous mobile sensing
  data.
\newblock {\em arXiv preprint arXiv:2007.09249}, 2020.

\bibitem{SADEGHIESHKEVARI2020106733}
S.~S. Eshkevari, T.~J. Matarazzo, and S.~N. Pakzad.
\newblock Bridge modal identification using acceleration measurements within
  moving vehicles.
\newblock {\em Mechanical Systems and Signal Processing}, 141:106733, 2020.

\bibitem{eshkevari2020simplified}
S.~S. Eshkevari, T.~J. Matarazzo, and S.~N. Pakzad.
\newblock Simplified vehicle--bridge interaction for medium to long-span
  bridges subject to random traffic load.
\newblock {\em Journal of Civil Structural Health Monitoring}, pages 1--15,
  2020.

\bibitem{eshkevari2019bridge}
S.~S. Eshkevari and S.~Pakzad.
\newblock Bridge structural identification using moving vehicle acceleration
  measurements.
\newblock In {\em Dynamics of Civil Structures, Volume 2}, pages 251--261.
  Springer, 2019.

\bibitem{fan2011vibration}
W.~Fan and P.~Qiao.
\newblock Vibration-based damage identification methods: a review and
  comparative study.
\newblock {\em Structural health monitoring}, 10(1):83--111, 2011.

\bibitem{farrar1996damage}
C.~R. Farrar and D.~Jauregui.
\newblock {\em Damage detection algorithms applied to experimental and
  numerical modal data from the I-40 bridge}.
\newblock Los Alamos National Laboratory, 1996.

\bibitem{feng2015citizen}
M.~Feng, Y.~Fukuda, M.~Mizuta, and E.~Ozer.
\newblock Citizen sensors for shm: Use of accelerometer data from smartphones.
\newblock {\em Sensors}, 15(2):2980--2998, 2015.

\bibitem{flandrin2004empirical}
P.~Flandrin, G.~Rilling, and P.~Goncalves.
\newblock Empirical mode decomposition as a filter bank.
\newblock {\em IEEE signal processing letters}, 11(2):112--114, 2004.

\bibitem{frangopol2001reliability}
D.~M. Frangopol, J.~S. Kong, and E.~S. Gharaibeh.
\newblock Reliability-based life-cycle management of highway bridges.
\newblock {\em Journal of computing in civil engineering}, 15(1):27--34, 2001.

\bibitem{frangopol1997life}
D.~M. Frangopol, K.-Y. Lin, and A.~C. Estes.
\newblock Life-cycle cost design of deteriorating structures.
\newblock {\em Journal of structural engineering}, 123(10):1390--1401, 1997.

\bibitem{furtado2014cost}
M.~N. Furtado and A.~A. Alipour.
\newblock Cost assessment of highway bridge network subjected to extreme
  seismic events.
\newblock {\em Transportation Research Record}, 2459(1):29--36, 2014.

\bibitem{gonzalez2012identification}
A.~Gonz{\'a}lez, E.~J. OBrien, and P.~McGetrick.
\newblock Identification of damping in a bridge using a moving instrumented
  vehicle.
\newblock {\em Journal of Sound and Vibration}, 331(18):4115--4131, 2012.

\bibitem{gonzalez2008understanding}
M.~C. Gonzalez, C.~A. Hidalgo, and A.-L. Barabasi.
\newblock Understanding individual human mobility patterns.
\newblock {\em nature}, 453(7196):779--782, 2008.

\bibitem{goupillaud1984cycle}
P.~Goupillaud, A.~Grossmann, and J.~Morlet.
\newblock Cycle-octave and related transforms in seismic signal analysis.
\newblock {\em Geoexploration}, 23(1):85--102, 1984.

\bibitem{guillemain1996characterization}
P.~Guillemain and R.~Kronland-Martinet.
\newblock Characterization of acoustic signals through continuous linear
  time-frequency representations.
\newblock {\em Proceedings of the IEEE}, 84(4):561--585, 1996.

\bibitem{guo2020semi}
K.~Guo and M.~J. Buehler.
\newblock A semi-supervised approach to architected materials design using
  graph neural networks.
\newblock {\em Extreme Mechanics Letters}, 41:101029, 2020.

\bibitem{hawelka2014geo}
B.~Hawelka, I.~Sitko, E.~Beinat, S.~Sobolevsky, P.~Kazakopoulos, and C.~Ratti.
\newblock {Geo-located Twitter as proxy for global mobility pattern}.
\newblock {\em Cartography and Geographic Information Science}, 41(3):260--271,
  2014.

\bibitem{he2016deep}
K.~He, X.~Zhang, S.~Ren, and J.~Sun.
\newblock Deep residual learning for image recognition.
\newblock In {\em Proceedings of the IEEE conference on computer vision and
  pattern recognition}, pages 770--778, 2016.

\bibitem{huang1998empirical}
N.~E. Huang, Z.~Shen, S.~R. Long, M.~C. Wu, H.~H. Shih, Q.~Zheng, N.-C. Yen,
  C.~C. Tung, and H.~H. Liu.
\newblock The empirical mode decomposition and the hilbert spectrum for
  nonlinear and non-stationary time series analysis.
\newblock {\em Proceedings of the Royal Society of London. Series A:
  Mathematical, Physical and Engineering Sciences}, 454(1971):903--995, 1998.

\bibitem{james1993natural}
G.~H. James~III, T.~G. Carne, and J.~P. Lauffer.
\newblock The natural excitation technique (next) for modal parameter
  extraction from operating wind turbines.
\newblock {\em NASA STI/Recon Technical Report N}, 93:28603, 1993.

\bibitem{jiang2017instantaneous}
Q.~Jiang and B.~W. Suter.
\newblock Instantaneous frequency estimation based on synchrosqueezing wavelet
  transform.
\newblock {\em Signal Processing}, 138:167--181, 2017.

\bibitem{jin2016damage}
C.~Jin, S.~Jang, X.~Sun, J.~Li, and R.~Christenson.
\newblock Damage detection of a highway bridge under severe temperature changes
  using extended kalman filter trained neural network.
\newblock {\em Journal of Civil Structural Health Monitoring}, 6(3):545--560,
  2016.

\bibitem{kim2007vibration}
J.-T. Kim, J.-H. Park, and B.-J. Lee.
\newblock Vibration-based damage monitoring in model plate-girder bridges under
  uncertain temperature conditions.
\newblock {\em Engineering Structures}, 29(7):1354--1365, 2007.

\bibitem{kong2003evaluation}
J.~S. Kong and D.~M. Frangopol.
\newblock Evaluation of expected life-cycle maintenance cost of deteriorating
  structures.
\newblock {\em Journal of Structural Engineering}, 129(5):682--691, 2003.

\bibitem{krizhevsky2012imagenet}
A.~Krizhevsky, I.~Sutskever, and G.~E. Hinton.
\newblock Imagenet classification with deep convolutional neural networks.
\newblock {\em Advances in neural information processing systems},
  25:1097--1105, 2012.

\bibitem{kuipers1999quaternions}
J.~B. Kuipers.
\newblock {\em Quaternions and rotation sequences: a primer with applications
  to orbits, aerospace, and virtual reality}.
\newblock Princeton university press, 1999.

\bibitem{lee2005neural}
J.~J. Lee, J.~W. Lee, J.~H. Yi, C.~B. Yun, and H.~Y. Jung.
\newblock Neural networks-based damage detection for bridges considering errors
  in baseline finite element models.
\newblock {\em Journal of Sound and Vibration}, 280(3-5):555--578, 2005.

\bibitem{leonard1992dynamic}
J.~J. Leonard, H.~F. Durrant-Whyte, and I.~J. Cox.
\newblock Dynamic map building for an autonomous mobile robot.
\newblock {\em The International Journal of Robotics Research}, 11(4):286--298,
  1992.

\bibitem{li2012reliability}
H.~Li, S.~Li, J.~Ou, and H.~Li.
\newblock Reliability assessment of cable-stayed bridges based on structural
  health monitoring techniques.
\newblock {\em Structure and Infrastructure Engineering}, 8(9):829--845, 2012.

\bibitem{Li2019}
X.~Li and C.~Ratti.
\newblock {Using Google Street View for street-level urban form analysis, a
  case study in Cambridge, Massachusetts}.
\newblock In L.~D'Acci, editor, {\em The Mathematics of Urban Morphology},
  pages 471--479. Springer International Publishing, 2019.

\bibitem{liang2018frequency}
Y.~Liang, D.~Li, G.~Song, and Q.~Feng.
\newblock Frequency co-integration-based damage detection for bridges under the
  influence of environmental temperature variation.
\newblock {\em Measurement}, 125:163--175, 2018.

\bibitem{lin2005use}
C.~Lin and Y.~Yang.
\newblock Use of a passing vehicle to scan the fundamental bridge frequencies:
  An experimental verification.
\newblock {\em Engineering Structures}, 27(13):1865--1878, 2005.

\bibitem{mallat1992characterization}
S.~Mallat and S.~Zhong.
\newblock Characterization of signals from multiscale edges.
\newblock {\em IEEE Transactions on Pattern Analysis \& Machine Intelligence},
  14(7):710--732, 1992.

\bibitem{marulanda2016modal}
J.~Marulanda, J.~M. Caicedo, and P.~Thomson.
\newblock Modal identification using mobile sensors under ambient excitation.
\newblock {\em Journal of Computing in Civil Engineering}, 31(2):04016051,
  2016.

\bibitem{marvasti2012nonuniform}
F.~Marvasti.
\newblock {\em Nonuniform sampling: theory and practice}.
\newblock Springer Science \& Business Media, 2012.

\bibitem{massaro2016car}
E.~Massaro, C.~Ahn, C.~Ratti, P.~Santi, R.~Stahlmann, A.~Lamprecht, M.~Roehder,
  and M.~Huber.
\newblock The car as an ambient sensing platform [point of view].
\newblock {\em Proceedings of the IEEE}, 105(1):3--7, 2016.

\bibitem{Massaro2017}
E.~Massaro, C.~Ahn, C.~Ratti, P.~Santi, R.~Stahlmann, A.~Lamprecht, M.~Roehder,
  and M.~Huber.
\newblock {The Car as an Ambient Sensing Platform}.
\newblock {\em Proceedings of the IEEE}, 105(1):3--7, 2017.

\bibitem{Massaro2019}
E.~Massaro, D.~Kondor, and C.~Ratti.
\newblock {Assessing the interplay between human mobility and mosquito borne
  diseases in urban environments}.
\newblock {\em Scientific Reports}, 9:16911, 2019.

\bibitem{matarazzo2016stride}
T.~J. Matarazzo and S.~N. Pakzad.
\newblock Stride for structural identification using expectation maximization:
  iterative output-only method for modal identification.
\newblock {\em Journal of Engineering Mechanics}, 142(4):04015109, 2016.

\bibitem{matarazzo2018scalable}
T.~J. Matarazzo and S.~N. Pakzad.
\newblock Scalable structural modal identification using dynamic sensor network
  data with stridex.
\newblock {\em Computer-Aided Civil and Infrastructure Engineering},
  33(1):4--20, 2018.

\bibitem{matarazzo2018crowdsensing}
T.~J. Matarazzo, P.~Santi, S.~N. Pakzad, K.~Carter, C.~Ratti, B.~Moaveni,
  C.~Osgood, and N.~Jacob.
\newblock Crowdsensing framework for monitoring bridge vibrations using moving
  smartphones.
\newblock {\em Proceedings of the IEEE}, 106(4):577--593, 2018.

\bibitem{mcaulay1986speech}
R.~McAulay and T.~Quatieri.
\newblock Speech analysis/synthesis based on a sinusoidal representation.
\newblock {\em IEEE Transactions on Acoustics, Speech, and Signal Processing},
  34(4):744--754, 1986.

\bibitem{mcgetrick2017implementation}
P.~McGetrick, D.~Hester, and S.~Taylor.
\newblock Implementation of a drive-by monitoring system for transport
  infrastructure utilising smartphone technology and gnss.
\newblock {\em Journal of Civil Structural Health Monitoring}, 7(2):175--189,
  2017.

\bibitem{mcgetrick2015experimental}
P.~J. McGetrick, C.-W. Kim, A.~Gonz{\'a}lez, and E.~J. Brien.
\newblock Experimental validation of a drive-by stiffness identification method
  for bridge monitoring.
\newblock {\em Structural Health Monitoring}, 14(4):317--331, 2015.

\bibitem{mei2020towards}
Q.~Mei, M.~G{\"u}l, and N.~Shirzad-Ghaleroudkhani.
\newblock Towards smart cities: crowdsensing-based monitoring of transportation
  infrastructure using in-traffic vehicles.
\newblock {\em Journal of Civil Structural Health Monitoring}, 10:653--665,
  2020.

\bibitem{milliken2002chassis}
W.~F. Milliken, D.~L. Milliken, and M.~Olley.
\newblock {\em Chassis design: principles and analysis}, volume 400.
\newblock Society of Automotive Engineers Warrendale, 2002.

\bibitem{mohan2008nericell}
P.~Mohan, V.~N. Padmanabhan, and R.~Ramjee.
\newblock Nericell: rich monitoring of road and traffic conditions using mobile
  smartphones.
\newblock In {\em Proceedings of the 6th ACM conference on Embedded network
  sensor systems}, pages 323--336, 2008.

\bibitem{mohseni2014simplified}
I.~Mohseni, A.~K.~A. Rashid, and J.~Kang.
\newblock A simplified method to estimate the fundamental frequency of skew
  continuous multicell box-girder bridges.
\newblock {\em Latin American Journal of Solids and Structures}, 11:649--658,
  2014.

\bibitem{nair2006time}
K.~K. Nair, A.~S. Kiremidjian, and K.~H. Law.
\newblock Time series-based damage detection and localization algorithm with
  application to the asce benchmark structure.
\newblock {\em Journal of Sound and Vibration}, 291(1-2):349--368, 2006.

\bibitem{ASCE2017}
A.~S. of~Civil~Engineers.
\newblock 2017 infrastructure report card on bridges.
\newblock \url{https://tinyurl.com/y5djfcnp}, March 2017.

\bibitem{OKeeffe12752}
K.~P. OKeeffe, A.~Anjomshoaa, S.~H. Strogatz, P.~Santi, and C.~Ratti.
\newblock Quantifying the sensing power of vehicle fleets.
\newblock {\em Proceedings of the National Academy of Sciences},
  116(26):12752--12757, 2019.

\bibitem{olley1934independent}
M.~Olley.
\newblock Independent wheel suspension—its whys and wherefores.
\newblock {\em SAE Transactions}, pages 73--81, 1934.

\bibitem{ozer2015citizen}
E.~Ozer, M.~Feng, and D.~Feng.
\newblock Citizen sensors for shm: Towards a crowdsourcing platform.
\newblock {\em Sensors}, 15(6):14591--14614, 2015.

\bibitem{pakzad2009statistical}
S.~N. Pakzad and G.~L. Fenves.
\newblock Statistical analysis of vibration modes of a suspension bridge using
  spatially dense wireless sensor network.
\newblock {\em Journal of structural engineering}, 135(7):863--872, 2009.

\bibitem{pakzad2008design}
S.~N. Pakzad, G.~L. Fenves, S.~Kim, and D.~E. Culler.
\newblock Design and implementation of scalable wireless sensor network for
  structural monitoring.
\newblock {\em Journal of infrastructure systems}, 14(1):89--101, 2008.

\bibitem{pandey1994damage}
A.~Pandey and M.~Biswas.
\newblock Damage detection in structures using changes in flexibility.
\newblock {\em Journal of sound and vibration}, 169(1):3--17, 1994.

\bibitem{peeters2001one}
B.~Peeters and G.~De~Roeck.
\newblock One-year monitoring of the z24-bridge: environmental effects versus
  damage events.
\newblock {\em Earthquake engineering \& structural dynamics}, 30(2):149--171,
  2001.

\bibitem{Pelletier2011}
M.-P. Pelletier, M.~Tr{\'{e}}panier, and C.~Morency.
\newblock {Smart card data use in public transit: A literature review}.
\newblock {\em Transportation Research Part C}, 19(4):557--568, 2011.

\bibitem{poncelet2007output}
F.~Poncelet, G.~Kerschen, J.-C. Golinval, and D.~Verhelst.
\newblock Output-only modal analysis using blind source separation techniques.
\newblock {\em Mechanical systems and signal processing}, 21(6):2335--2358,
  2007.

\bibitem{rabiei2013rainfall}
E.~Rabiei, U.~Haberlandt, M.~Sester, and D.~Fitzner.
\newblock Rainfall estimation using moving cars as rain gauges--laboratory
  experiments.
\newblock {\em Hydrology and Earth System Sciences}, 17(11):4701--4712, 2013.

\bibitem{raissi2018hidden}
M.~Raissi and G.~E. Karniadakis.
\newblock Hidden physics models: Machine learning of nonlinear partial
  differential equations.
\newblock {\em Journal of Computational Physics}, 357:125--141, 2018.

\bibitem{ralbovsky2010frequency}
M.~Ralbovsky, S.~Deix, and R.~Flesch.
\newblock Frequency changes in frequency-based damage identification.
\newblock {\em Structure and Infrastructure Engineering}, 6(5):611--619, 2010.

\bibitem{rioul1991wavelets}
O.~Rioul and M.~Vetterli.
\newblock Wavelets and signal processing.
\newblock {\em IEEE signal processing magazine}, 8(ARTICLE):14--38, 1991.

\bibitem{ruzzene1997natural}
M.~Ruzzene, A.~Fasana, L.~Garibaldi, and B.~Piombo.
\newblock Natural frequencies and dampings identification using wavelet
  transform: application to real data.
\newblock {\em Mechanical systems and signal processing}, 11(2):207--218, 1997.

\bibitem{sahin2003quantification}
M.~Sahin and R.~Shenoi.
\newblock Quantification and localisation of damage in beam-like structures by
  using artificial neural networks with experimental validation.
\newblock {\em Engineering structures}, 25(14):1785--1802, 2003.

\bibitem{schlune2009improved}
H.~Schlune, M.~Plos, and K.~Gylltoft.
\newblock Improved bridge evaluation through finite element model updating
  using static and dynamic measurements.
\newblock {\em Engineering structures}, 31(7):1477--1485, 2009.

\bibitem{Schneider2013}
C.~M. Schneider, V.~Belik, T.~Couronn{\'{e}}, Z.~Smoreda, and M.~C.
  Gonz{\'{a}}lez.
\newblock {Unravelling daily human mobility motifs.}
\newblock {\em Journal of the Royal Society, Interface}, 10(84):20130246, 2013.

\bibitem{shirzad2020frequency}
N.~Shirzad-Ghaleroudkhani, Q.~Mei, and M.~G{\"u}l.
\newblock Frequency identification of bridges using smartphones on vehicles
  with variable features.
\newblock {\em Journal of Bridge Engineering}, 25(7):04020041, 2020.

\bibitem{shoemake1985animating}
K.~Shoemake.
\newblock Animating rotation with quaternion curves.
\newblock In {\em Proceedings of the 12th annual conference on Computer
  graphics and interactive techniques}, pages 245--254, 1985.

\bibitem{siringoringo2012estimating}
D.~M. Siringoringo and Y.~Fujino.
\newblock Estimating bridge fundamental frequency from vibration response of
  instrumented passing vehicle: analytical and experimental study.
\newblock {\em Advances in Structural Engineering}, 15(3):417--433, 2012.

\bibitem{sitton2020bridge}
J.~D. Sitton, D.~Rajan, and B.~A. Story.
\newblock Bridge frequency estimation strategies using smartphones.
\newblock {\em Journal of Civil Structural Health Monitoring}, pages 1--14,
  2020.

\bibitem{sitton2020frequency}
J.~D. Sitton, Y.~Zeinali, D.~Rajan, and B.~A. Story.
\newblock Frequency estimation on two-span continuous bridges using dynamic
  responses of passing vehicles.
\newblock {\em Journal of Engineering Mechanics}, 146(1):04019115, 2020.

\bibitem{smith2016studies}
I.~F. Smith.
\newblock Studies of sensor data interpretation for asset management of the
  built environment.
\newblock {\em Frontiers in Built Environment}, 2:8, 2016.

\bibitem{bbva}
S.~Sobolevsky, I.~Sitko, R.~T. {Des Combes}, B.~Hawelka, J.~M. Arias, and
  C.~Ratti.
\newblock {Cities through the Prism of People's Spending Behavior}.
\newblock {\em PLoS ONE}, 11(2):e0146291, 2016.

\bibitem{staszewski1997identification}
W.~Staszewski.
\newblock Identification of damping in mdof systems using time-scale
  decomposition.
\newblock {\em Journal of sound and vibration}, 203(2):283--305, 1997.

\bibitem{taysi2012routing}
Z.~C. Taysi and A.~G. Yavuz.
\newblock Routing protocols for geonet: A survey.
\newblock {\em IEEE Transactions on Intelligent Transportation Systems},
  13(2):939--954, 2012.

\bibitem{thakur2013synchrosqueezing}
G.~Thakur, E.~Brevdo, N.~S. Fu{\v{c}}kar, and H.-T. Wu.
\newblock The synchrosqueezing algorithm for time-varying spectral analysis:
  Robustness properties and new paleoclimate applications.
\newblock {\em Signal Processing}, 93(5):1079--1094, 2013.

\bibitem{vaswani2017attention}
A.~Vaswani, N.~Shazeer, N.~Parmar, J.~Uszkoreit, L.~Jones, A.~N. Gomez,
  {\L}.~Kaiser, and I.~Polosukhin.
\newblock Attention is all you need.
\newblock In {\em Advances in neural information processing systems}, pages
  5998--6008, 2017.

\bibitem{wahab1999damage}
M.~A. Wahab and G.~De~Roeck.
\newblock Damage detection in bridges using modal curvatures: application to a
  real damage scenario.
\newblock {\em Journal of Sound and vibration}, 226(2):217--235, 1999.

\bibitem{wang2013matching}
S.~Wang, X.~Chen, G.~Cai, B.~Chen, X.~Li, and Z.~He.
\newblock Matching demodulation transform and synchrosqueezing in
  time-frequency analysis.
\newblock {\em IEEE Transactions on Signal Processing}, 62(1):69--84, 2013.

\bibitem{xiao2015multiscale}
X.~Xiao, Y.~L. Xu, and Q.~Zhu.
\newblock Multiscale modeling and model updating of a cable-stayed bridge. ii:
  Model updating using modal frequencies and influence lines.
\newblock {\em Journal of Bridge Engineering}, 20(10):04014113, 2015.

\bibitem{yang2004hilbert}
J.~N. Yang, Y.~Lei, S.~Lin, and N.~Huang.
\newblock Hilbert-huang based approach for structural damage detection.
\newblock {\em Journal of engineering mechanics}, 130(1):85--95, 2004.

\bibitem{yang2009extracting}
Y.~Yang and K.~Chang.
\newblock Extracting the bridge frequencies indirectly from a passing vehicle:
  Parametric study.
\newblock {\em Engineering Structures}, 31(10):2448--2459, 2009.

\bibitem{yang2013frequency}
Y.~Yang, M.~Cheng, and K.~Chang.
\newblock Frequency variation in vehicle--bridge interaction systems.
\newblock {\em International Journal of Structural Stability and Dynamics},
  13(02):1350019, 2013.

\bibitem{yang1995vehicle}
Y.-B. Yang and B.-H. Lin.
\newblock Vehicle-bridge interaction analysis by dynamic condensation method.
\newblock {\em Journal of Structural Engineering}, 121(11):1636--1643, 1995.

\bibitem{yang2004extracting}
Y.-B. Yang, C.~Lin, and J.~Yau.
\newblock Extracting bridge frequencies from the dynamic response of a passing
  vehicle.
\newblock {\em Journal of Sound and Vibration}, 272(3-5):471--493, 2004.

\bibitem{yang2020vehicle}
Y.-B. Yang, J.~P. Yang, B.~Zhang, and Y.~Wu.
\newblock {\em Vehicle Scanning Method for Bridges}.
\newblock Wiley Online Library, 2020.

\bibitem{yang1997vehicle}
Y.-B. Yang and J.-D. Yau.
\newblock Vehicle-bridge interaction element for dynamic analysis.
\newblock {\em Journal of Structural Engineering}, 123(11):1512--1518, 1997.

\bibitem{yao2012autoregressive}
R.~Yao and S.~N. Pakzad.
\newblock Autoregressive statistical pattern recognition algorithms for damage
  detection in civil structures.
\newblock {\em Mechanical Systems and Signal Processing}, 31:355--368, 2012.

\bibitem{zabini2016random}
F.~Zabini, A.~Calisti, D.~Dardari, and A.~Conti.
\newblock Random sampling via sensor networks: Estimation accuracy vs. energy
  consumption.
\newblock In {\em 2016 24th European Signal Processing Conference (EUSIPCO)},
  pages 130--134. IEEE, 2016.

\bibitem{zabini2016inhomogeneous}
F.~Zabini and A.~Conti.
\newblock Inhomogeneous poisson sampling of finite-energy signals with
  uncertainties in $r^d$.
\newblock {\em IEEE Transactions on Signal Processing}, 64(18):4679--4694,
  2016.

\bibitem{zhang2012damage}
Y.~Zhang, L.~Wang, and Z.~Xiang.
\newblock Damage detection by mode shape squares extracted from a passing
  vehicle.
\newblock {\em Journal of Sound and Vibration}, 331(2):291--307, 2012.

\bibitem{zhong2005phase}
J.~Zhong and J.~Weng.
\newblock Phase retrieval of optical fringe patterns from the ridge of a
  wavelet transform.
\newblock {\em Optics letters}, 30(19):2560--2562, 2005.

\end{thebibliography}

\end{document}